\documentclass{emulateapj}

\usepackage{graphicx}
\usepackage[hypertex]{hyperref}

\def \eg {e.g.}
\def \ie {i.e.}
\def\spose#1{\hbox to 0pt{#1\hss}}
\def\ltsim{$\mathrel{\spose{\lower 3pt\hbox{$\sim$}}
        \raise 2.0pt\hbox{$<$}}$\thinspace}
\def\gtsim{$\mathrel{\spose{\lower 3pt\hbox{$\sim$}}
        \raise 2.0pt\hbox{$>$}}$\thinspace}
\newcommand{\thin }{\thinspace}

\newcommand{\etc}{etc.}

\newcommand{\lk }{${\rm L_K}$}

\newcommand{\msun }{${\rm M_{\odot}}$}
\newcommand{\lsun }{${\rm L_{\odot}}$}

\newcommand{\ergps }{${\rm erg\ s^{-1}}$}
\newcommand{\mvir}{${\rm M_{vir}}$}
\newcommand{\rvir}{${\rm R_{vir}}$}
\newcommand{\mfive}{${\rm M_{500}}$}
\newcommand{\rfive}{${\rm R_{500}}$}
\newcommand{\rtwo}{${\rm R_{200}}$}
\newcommand{\mtwo}{${\rm M_{200}}$}
\newcommand{\rtwelve}{${\rm R_{1250}}$}
\newcommand{\rtwentyfive}{${\rm R_{2500}}$}
\newcommand{\mtwentyfive}{${\rm M_{2500}}$}
\newcommand{\cvir}{${\rm c_{vir}}$}

\newcommand{\reff}{${\rm R_{e}}$}
\newcommand{\src }{NGC\thin 720}
\newcommand{\zfe }{${\rm Z_{Fe}}$}

\newcommand{\suzaku}{{\em Suzaku}}
\newcommand{\chandra }{{\em Chandra}}

\newcommand{\xspec }{{\em Xspec}}

\newcommand{\lstar}{${\rm L_*}$}

\newcommand{\minuit}{MINUIT}
\newcommand{\ciao }{{\em CIAO}}
\newcommand{\caldb }{{\em Caldb}}
\newcommand{\heasoft }{{\em Heasoft}}
\newcommand{\ned}{{\em{NED}}}

\newcommand{\xmm }{{\em XMM}}

\newcommand{\mbh} {${\rm M_{BH}}$}

\newcommand{\lx }{${\rm L_X}$}
\newcommand{\fb}{${\rm f_b}$}
\newcommand{\fbvir}{${\rm f_{b,vir}}$}
\newcommand{\fbtwentyfive}{${\rm f_{b,2500}}$}
\newcommand{\fbfive}{${\rm f_{b,500}}$}
\newcommand{\fbtwo}{${\rm f_{b,200}}$}

\newcommand{\fgas}{${\rm f_{g}}$}
\newcommand{\fg}{\fgas}
\newcommand{\kfive}{${\rm K_{500}}$}

\newcommand{\lb }{${\rm L_B}$}

\newcommand{\dtwentyfive}{${\rm D_{25}}$}
\newcommand{\twomass}{2MASS}
\newcommand{\sigmac}{$\sigma_*$}

\defcitealias{humphrey06a}{H06}
\defcitealias{humphrey08a}{H08}

\begin{document}
\title{A census of baryons and dark matter in an isolated, Milky Way-sized elliptical galaxy}
\author{Philip J. Humphrey\altaffilmark{1}, David A. Buote\altaffilmark{1}, Claude R. Canizares\altaffilmark{2}, Andrew C. Fabian\altaffilmark{3}, Jon M. Miller\altaffilmark{4} }
\altaffiltext{1}{Department of Physics and Astronomy, University of California at Irvine, 4129 Frederick Reines Hall, Irvine, CA 92697}
\altaffiltext{2}{Department of Physics and Kavli Institute for Astrophysics and Space Research, Massachusetts 
Institute of Technology, Cambridge, MA 02139}
\altaffiltext{3}{Institute of Astronomy, Madingley Road, Cambridge CB3 0HA, UK}
\altaffiltext{4}{Department of Astronomy, University of Michigan, Ann Arbor, Michigan 48109}
\begin{abstract}
We present a study of the dark and luminous matter in the isolated elliptical galaxy \src,
based on deep X-ray observations made with the \chandra\ and \suzaku\ observatories.  
The gas properties are reliably measured almost to \rtwentyfive, allowing us to place
good constraints on the enclosed mass and baryon fraction (\fb) within this radius
(\mtwentyfive=$1.6\pm0.2 \times 10^{12}$\msun, \fbtwentyfive=$0.10\pm0.01$; 
systematic errors are typically \ltsim 20\%). 
{The data indicate that the hot gas is close to hydrostatic,
which is supported by good agreement with a kinematical analysis of the dwarf 
satellite galaxies. We confirm at high significance
($\sim$20-$\sigma$) the presence of a dark matter (DM) halo. Assuming an NFW DM
profile, our physical model for the gas distribution enables 
us to obtain meaningful constraints at scales larger than \rtwentyfive, revealing
that most of the baryons are in the hot gas.
We find that 
\fb\ within the virial radius is consistent with the Cosmological
value, confirming theoretical predictions that a 
$\sim$Milky Way-mass (\mvir=$3.1^{+0.4}_{-0.3} \times 10^{12}$\msun) galaxy can
sustain a massive, quasi-hydrostatic gas halo.}
While \fb\ is higher than the cold (cool gas plus stars) 
baryon fraction typically measured in similar-mass spiral galaxies, both the gas 
fraction (\fg)
and \fb\  in \src\ are consistent with an 
extrapolation of the trends with mass seen in massive galaxy groups and clusters.
{After correcting for \fg, the entropy profile is close to the self-similar prediction
of gravitational structure formation simulations, as observed in massive galaxy clusters.}
Finally, we find a strong heavy metal abundance gradient in the interstellar medium, 
qualitatively similar to those observed in massive galaxy groups.
\end{abstract}
\section{Introduction} \label{sect_introduction}
In low-redshift disk galaxies of size comparable to the Milky Way, the
fraction of cold baryons (cool gas and stars) relative to the total mass is
typically found to be significantly less than the cosmological baryon
fraction  \citep[\eg][]{fukugita98a,mcgaugh10a}, which is measured
to be 0.17 \citep{dunkley09a}.
This has proven a challenge for standard models
of galaxy formation which, in fact, over-predict the observed baryon
content by as much as a factor $\sim$2 \citep[\eg][]{benson03a}.
Solutions to this problem, for example strong heating of the interstellar medium (ISM) by
active galactic nucleus (AGN) feedback \citep[\eg][]{kauffmann99a} or less efficient cooling
\citep[\eg][]{maller04a}, ultimately involve many of the baryons
remaining in a difficult to detect hot phase, which may exist as an
extended halo around the galaxy. Models suggest that as many as
half the ``missing baryons'' in the local universe could be in
diffuse, pressure-supported halos around normal galaxies 
\citep[\eg][]{white91a,maller04a,keres05a,fukugita06a,sommerlarsen06a}.
Such a halo around the Milky Way could help explain the 
properties of local OVI absorbing clouds \citep{sembach03a},
the confinement of the Magellanic stream \citep{moore94a},
gas stripping from dwarf satellites \citep{blitz00a} and provide
a reservoir for the ongoing gas accretion proposed to solve the ``G-dwarf
problem'' \citep[\eg][]{sommerlarsen03a}.

Direct observational evidence for such extended halos
around disk galaxies remains controversial. While zero redshift
absorption lines in the X-ray spectra of distant quasars 
clearly indicates diffuse, hot ($\sim 10^6$K) gas in the vicinity
of the Milky Way 
\citep[\eg][]{nicastro02a,fang02a,rasmussen03a,fang06a,bregman07b,
buote09a,fang10a}, the distribution of this gas along the sight-line
remains unclear \citep[\eg][]{yao05a,yao08a}. Other tracers
of the putative hot baryons suggest the gas may be 
confined to the disk \citep[][although see \citealt{joudaki10a}]{anderson10a,henley10a},
but the results are highly model-dependent. Attempts to 
detect gaseous halos around external disk galaxies have,
similarly, proven inconclusive. 
While X-ray emission from diffuse, hot gas 
associated with star-formation in the disk has been observed 
\citep[\eg][]{fabbiano01a,strickland04a}, similar emission from an extended, hot
corona has yet to be securely detected 
\citep{benson00a,rasmussen09a}.

In contrast to disk galaxies, diffuse X-ray emission from 
massive galaxy groups and clusters has been detected out to scales
as large as $\sim$\rtwelve--\rfive\footnote{We define $R_\Delta$ as the geometrical
radius within which the mean mass density of the system
is $\Delta$ times the critical density
of the Universe.}, allowing tight constraints to be placed
on the total hot gas content 
\citep[\eg][]{gastaldello07a,sun08a,vikhlinin06b,allen08a,pratt09a}. While the most
massive clusters appear to be close to baryonic closure
(\ie\ the baryon fraction, \fb, at large radii asymptotes to the 
to the Cosmological value), \fb\ at lower masses 
may be significantly smaller \citep{gastaldello07a,pratt09a,giodini09a,dai09a}.
The gas entropy profiles of groups are
systematically enhanced over the predictions of self-similar formation
models, suggesting that 
non-gravitational processes (\eg\ feedback during halo assembly) 
are responsible
for ejecting baryons from the central regions of the halo
\citep[\eg][]{ponman99a,pratt10a,mccarthy10a}. Whether these baryons are completely
ejected from the potential well is, however, unclear;
extrapolating to \rvir, for example, \citet{gastaldello07a} found 
\fb\ close to the Cosmological value for their group sample.

Giant elliptical galaxies occupy an important niche between Milky-Way
sized disk galaxies and galaxy groups, offering a chance to connect these two
mass regimes. As the likely end products of spiral galaxy merging, their 
baryon content may also provide important constraints on how the baryons
are distributed in their progenitors. Unlike disk galaxies, {\em most} of 
their gas is typically in the form of an extended, hot halo, 
the X-ray emission from which is often detected out at least to tens of kpc.
While early-type galaxies therefore provide an ideal opportunity to take
a census of the baryon content in normal galaxies, surprisingly little
is known about their \fb. Based on weak lensing studies, the {\em stellar}
mass fraction appears to be small 
\citep[\ltsim 0.03:][]{hoekstra05a,gavazzi07a},
but for massive galaxies, most of the baryons are expected to
be in the hot halo \citep[\eg][]{keres05a}.  In their hydrostatic 
X-ray mass analysis of three isolated elliptical galaxies (and 
four low-mass groups),  \citet[][hereafter \citetalias{humphrey06a}]{humphrey06a} 
were only able to obtain interesting constraints on the total baryon content
by employing restrictive priors. 

\citet{mathews05a} showed that the 
X-ray luminosity (\lx) of the most X-ray bright early-type galaxies is consistent with their
being at the centre of massive, baryonically closed groups. However, there is 
observed to be a a huge range of \lx\ at 
fixed optical luminosity \citep{canizares87a,osullivan01a,ellis06a}, suggesting 
that \lx\ is determined by both the virial mass of the 
halo in which galaxies lie and their history of feedback \citep{mathews06a}.
In some cases, particularly in the lowest-mass systems, feedback appears to have been
strong enough to denude a galaxy largely of its gas \citep{david06a}. 
Still, it is difficult to map galaxies from the optical {\em versus} X-ray luminosity
plane onto \fb\ directly, in large part since the gas emissivity depends on the square of its 
density and gas density profiles are not self-similar \citep[\eg\ 
\citetalias{humphrey06a};][]{gastaldello07a,sun08a}.
This means that \lx, which is typically measured within only a small fraction of the 
virial radius (\rvir), is a poor tracer of the overall gas mass. 
The detailed spatially resolved studies needed to trace the gas distribution to large
enough radii to measure \fb\ more directly 
have largely been restricted to the most X-ray luminous systems, which tend to 
lie at the centres of massive groups. Thus, whether early-type galaxies hosted in Milky-Way sized
halos can actually maintain a hot halo with a mass comparable to the predictions of 
disk galaxy formation models remains an open question.

To begin to address this question, in this paper we present a detailed X-ray study
of the isolated elliptical galaxy NGC\thin 720, using deep \chandra\ and \suzaku\
observations (see Table~\ref{table_observations}). 
The complementary characteristics of these satellites,
\ie\ the excellent spatial resolution of \chandra\ (enabling the inner parts of the X-ray
halo to be studied in detail) and the low, stable background of \suzaku\ (which is helpful
for studying the low surface-brightness outer parts of the halo), make a combination
of these data ideal for studying the hot gas properties over as wide a radial range as
possible. \src\ is a well-studied galaxy with a modest \lx; 
for its optical luminosity, it lies roughly in the middle of the scatter in the 
\lx-\lb\ relation \citep{osullivan01a}. It has occasionally been labelled
a galaxy group \citep[\eg][]{garcia93,osmond04a}, but within \rfive\ the system comprises only 
one \lstar\ galaxy (NGC\thin 720 itself) and dwarf companions, the brightest of 
which is three magnitudes fainter than the central galaxy \citep{dressler86a,brough06a}. 
The total virial mass of the system, derived both from hydrostatic X-ray modelling
and from the kinematics of the satellites, is a few times $10^{12}$\msun\
(\citetalias{humphrey06a}; \citealt{brough06a}), making it only a few times more
massive than the Milky Way, and therefore the ideal system for our purposes. Past 
studies, however, have not provided good constraints on the overall baryon content
of the system \citepalias{humphrey06a}.

The X-ray emission from the hot gas in \src, which can be traced out to scales of 
at least $\sim$80~kpc \citepalias{humphrey06a}, has a relaxed morphology
\citep{buote94,buote02b}, suggesting that hydrostatic mass modelling
techniques can be reliably used to infer the total gravitating mass \citep[\eg][]{buote95a}. 
This is further supported by the lack of significant radio emission \citep{fabbiano87a},
indicating that there is not currently an AGN that might be stirring up the ISM.
For relaxed systems, hydrostatic equilibrium is believed to be an excellent
approximation, with non-thermal effects expected to contribute no more than
$\sim$20 per cent of the total pressure \citep[\eg][]{nagai07a,piffaretti08a,fang09a}.
The overall agreement between the masses inferred from hydrostatic methods 
and those found by gravitational lensing and stellar dynamical modelling or the 
predictions of stellar population synthesis models generally support this 
picture \citep[\eg][]{mahdavi08a,churazov08a,humphrey09d}. Modest
discrepancies have been reported in a few galaxies 
between hydrostatic X-ray masses and stellar dynamics results \citep[\eg][]{romanowsky09a,gebhardt09a,shen10a}, 
but these may simply reflect systematic uncertainties in the stellar dynamical modelling 
\citep[\eg][]{gavazzi05a,thomas07b}, or the dynamics of the hot gas at the very
smallest scales \citep{brighenti09a}. 

We adopted a nominal distance of 25.7~Mpc for \src, based on the I-band SBF distance
modulus estimate of \citet{tonry01}, which was corrected to account for recent revisions to
the Cepheid zero-point (see \citetalias{humphrey06a}). At this distance, 1\arcsec\ corresponds to 123~pc.
 We assumed a flat cosmology with $H_0 =70 {\rm km\ s^{-1}}$
and $\Omega_\Lambda=0.7$. We adopted $R_{102}$ as the virial radius (\rvir), based on the approximation
of \citet{bryan98a} for the redshift of \src.
Unless otherwise stated, all error-bars represent 1-$\sigma$ confidence limits (which, for our 
Bayesian analysis, implies the marginalized region of parameter space within which the integrated 
probability is 68\%).

\section{Data analysis}
\subsection{Chandra}
\begin{deluxetable}{llr}
\tablecaption{Observation summary\label{table_observations}}
\tablehead{
\colhead{ObsID} & \colhead{Start Date} & \colhead{Exposure (ks)}
}
\tablewidth{3in}
\startdata
\multicolumn{3}{c}{Chandra}\\
7062 & 2006 Oct 9 & 23 \\
7372 & 2006 Aug 6 & 49 \\
8448 & 2006 Oct 12 & 8\\
8449 & 2006 Oct 12 & 19\\ \hline
\multicolumn{3}{c}{Suzaku}\\ 
800009010 & 2005 Dec 30 & 177 
\enddata
\tablecomments{Details of the observations used in the present analysis. For each
dataset we quote the observation identification number 
(ObsID), the start date and the exposure time, after having
removed periods of background ``flaring'' (Exposure). An additional, shallow \chandra\
observation
(ObsID 492) was excluded due to its elevated background level.}
\end{deluxetable}
\begin{figure*}
\centering
\includegraphics[width=6.0in]{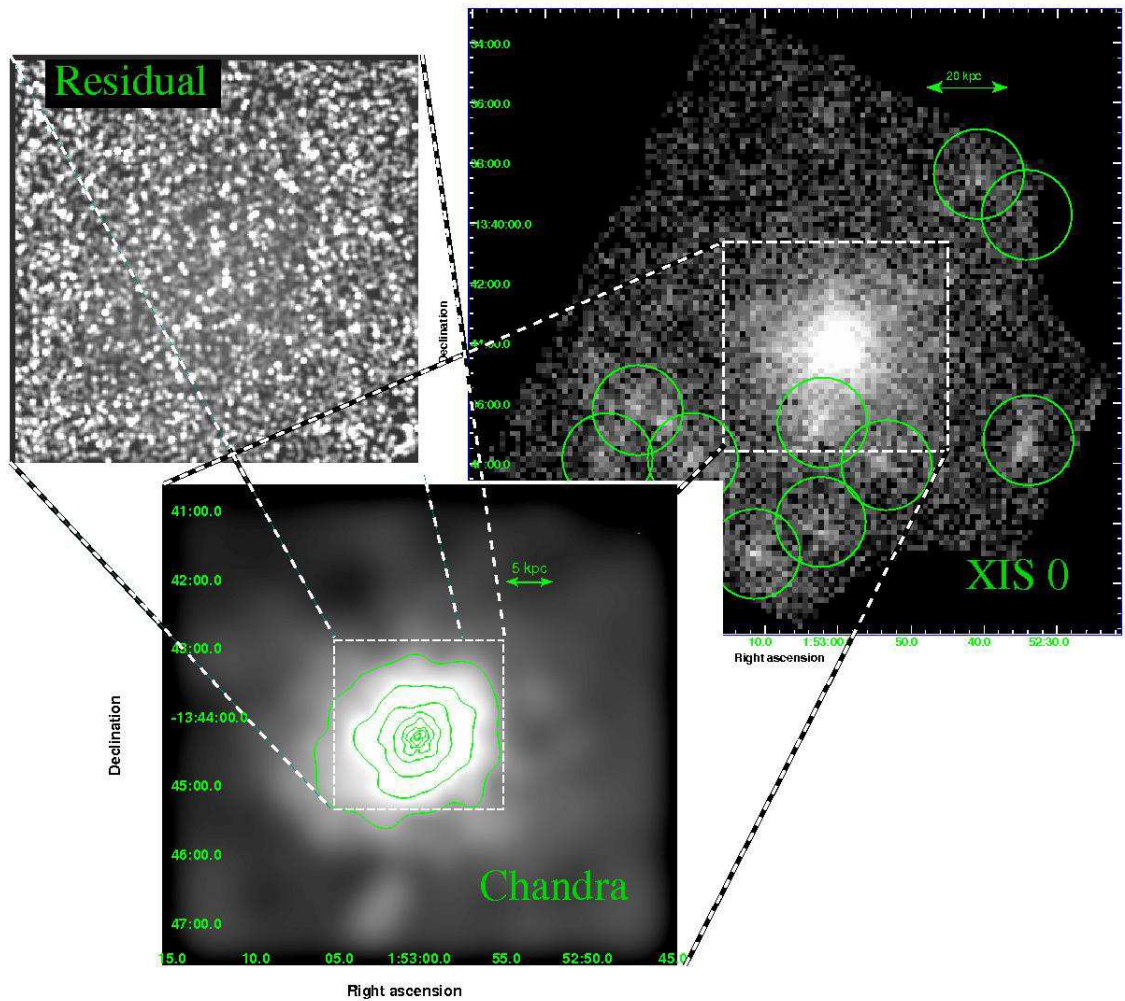}
\caption{ {\em Top right:} \suzaku\ XIS0 image in the 0.5--7.0~keV band, excluding data 
in the vicinity of the calibration sources. Circular exclusion regions (to mitigate bright
point-source contamination) are overlaid. {\em Bottom:}
Smoothed, point-source subtracted \chandra\ X-ray image 
in the 0.5--7.0~keV band. This image covers only the central part of the 
ACIS-S3 chip, where the count-rate is sufficiently high for the structure of the X-ray
emission to be clearly discerned. The image
contrast and smoothing scales were arbitrarily adjusted to bring out key features;
the smoothing scale varies from $\sim$1\arcsec\ in the central part of the images to 
$\sim$1\arcmin\ in the outer regions. Logarithmically spaced contours are overlaid
to guide the eye. {\em Top left:} ``residual significance'' image 
(see text) of the centre of the galaxy, indicating deviations from a smooth model
fit to the X-ray isophotes. Although the isophotes are slightly elliptical,
the galaxy appears relaxed and symmetric at all accessible scales given its distance.\label{fig_images}}
\end{figure*}
\subsubsection{Data reduction}
The region of sky containing  NGC\thin 720 has been imaged by the \chandra\ ACIS instrument
in the ACIS-S configuration on 4 separate occasions (a shallow fifth observation was 
excluded from our analysis due to its enhanced background level), 
as listed in Table~\ref{table_observations}.
We processed each dataset independently,
using the CIAO~4.1 and \heasoft\ 6.8
software suites, in conjunction with the \chandra\ calibration database (\caldb) version 4.1.2.
To ensure up-to-date calibration, all data were
reprocessed from the ``level 1'' events files, following the standard
\chandra\ data-reduction threads\footnote{{http://cxc.harvard.edu/ciao/threads/index.html}}.
We applied the standard correction to take account of the time-dependent gain-drift
and charge transfer inefficiency,
as implemented in the \ciao\ tools. To identify periods of enhanced
background (which can seriously degrade the signal-to-noise, S/N)
we accumulated background lightcurves for each dataset from
low surface-brightness regions of the active chips, excluding obvious
point-sources. Such periods of background ``flaring'' were 
identified by eye and
excised. The final exposure times are listed in Table~\ref{table_observations}.
To combine the individual datasets, we merged the final data products (images, spectra,
spectral response files, \etc) using the procedure outlined in 
\citet{humphrey08a}. This involves correcting for relative astrometric errors by 
matching the positions of detected point sources (allowing for the possibility
that some fraction of the sources are transient).
Point sources were identified in each observation by applying the 
\ciao\ {\tt wavdetect} task to the full-resolution 0.5--7.0~keV band image,
supplying the exposure-map (computed at an energy of 1.7~keV) 
to minimize spurious detections at the image boundaries.
The detection threshold was set to $10^{-6}$, corresponding to \ltsim 1 spurious source
detections per chip.
A final source list was obtained by repeating the source detection procedure on the 
final, merged image. All detected sources were confirmed by visual inspection, 
and, for each,
appropriate elliptical regions containing approximately 99\%\ of its photons were generated.
We detected 58 point sources within the \dtwentyfive\  ellipse \citep{devaucouleurs91},
contributing $\sim$20\%\ of the total 0.5--7.0~keV photons in that region.

\subsubsection{\chandra\ Image}
We examined the image for evidence of
morphological disturbances that may complicate deprojection, or possibly suggest
perturbation from hydrostatic equilibrium.
In Fig~\ref{fig_images} we show  smoothed, flat-fielded \chandra\ image, having removed 
the point-sources 
with the algorithm outlined in \citet{fang09a}.
The data were then flat-fielded with the 
exposure map and smoothed with a Gaussian kernel. Since the S/N of the image varies strongly
with off-axis angle, the width of the Gaussian kernel was varied
with distance according to an arbitrary power law, ranging from $\sim$1\arcsec\ in the centre
of the image to $\sim$1\arcmin\ at the edge of the field.

The image is smooth and slightly 
elliptical, showing no clear evidence of disturbances or asymmetries. 
This is consistent with our findings based on a shallower \chandra\ observation
\citep{buote02b}.
To search for more subtle structure in the image,
we used dedicated software, built around the \minuit\ library\footnote{http://lcgapp.cern.ch/project/cls/work-packages/mathlibs/minuit/index.html}, to fit an elliptical beta model
(with constant ellipticity) to the central $\sim$2.5\arcmin\ wide portion of the unsmoothed image, 
(correcting for exposure variations with the exposure map)\footnote{We obtained a best-fitting major-axis
core radius of 2.6\arcsec, $\beta=0.38$ and axis ratio of 0.85, consistent
with \citet{buote02b}}.  We show in Fig~\ref{fig_images} 
an indication of deviations from this simple model by plotting (data-model)$^2$/model, which 
corresponds to the $\chi^2$ residual in each pixel. To bring out structure, we smoothed this 
image with a Gaussian kernel of width 1.5\arcsec\ (3 pixels). 
As is clear from this ``residuals significance'' image, there is 
no evidence of any coherent residual features. This is unsurprising given the lack of a currently
active central black hole, as suggested by the lack a significant radio emission
\citep{fabbiano87a}, nor a central, bright X-ray source.
\subsubsection{Spectral analysis} \label{sect_chandra_spectrum}
\begin{figure*}
\centering
\includegraphics[width=6.0in]{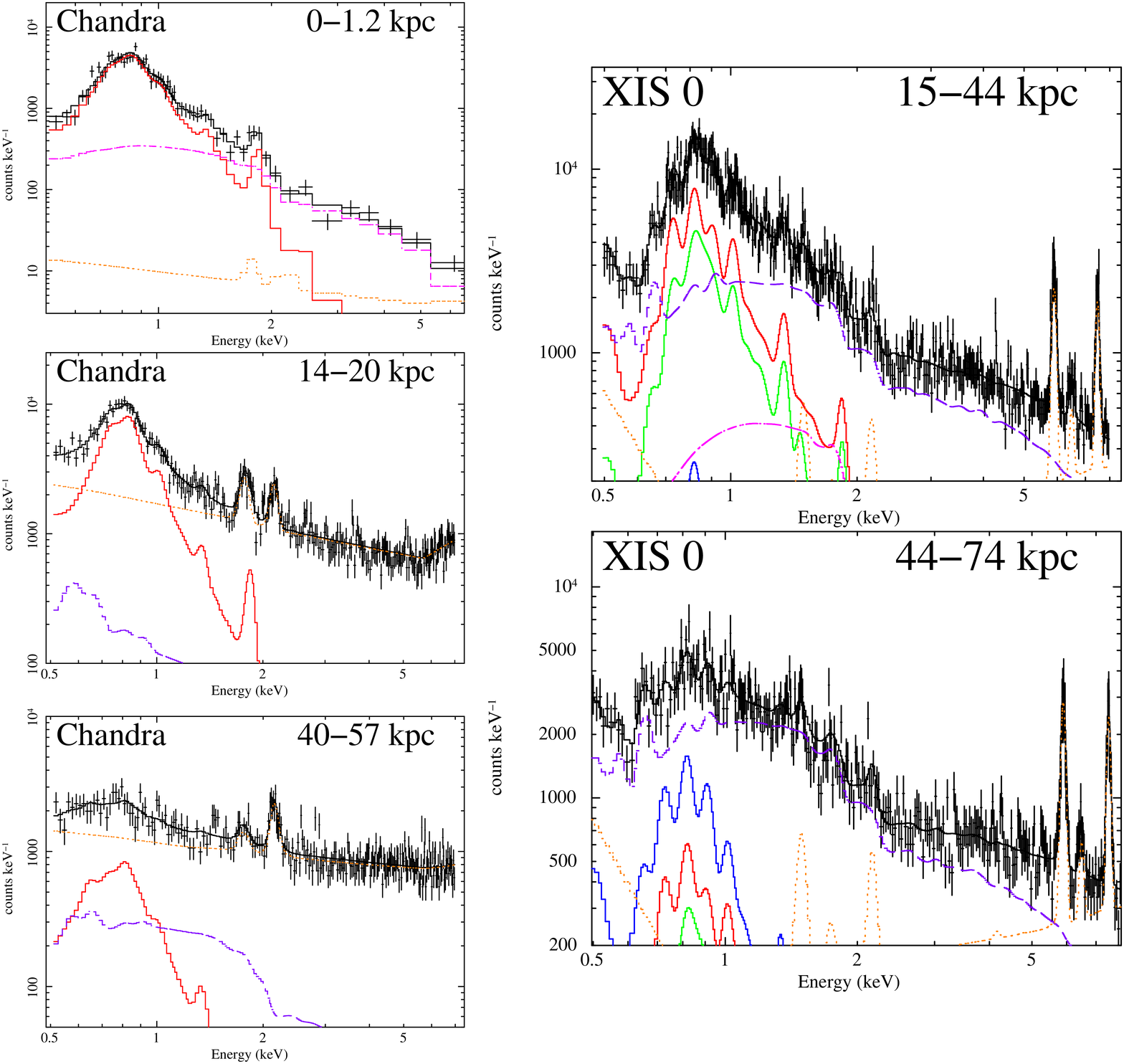}
\caption{Representative \chandra\ and \suzaku\ XIS0 spectra for \src, shown without background
subtraction. In addition to the data, we show the best-fitting model, folded through the instrumental
response (solid black line), along with the decomposition of this model into its various
components. For the \chandra\ data, we show the hot gas contribution (solid red line), the 
composite emission from X-ray binaries (dash-dot magenta line), the instrumental background
(dotted orange) and the cosmic X-ray background (dashed purple line). For the XIS0 data, we show the 
hot gas emission from each annulus as solid lines (green from the central annulus, red for the 
second annulus and blue for the third), as well as the X-ray binary component, instrumental
background and sky background (using the same colour scheme as for \chandra). For \chandra, the 
background is dominated by the instrumental component, 
whereas for \suzaku, which has a lower instrumental background but is less able to resolve
the cosmic component into individual point sources, the cosmic component dominates. In all cases,
emission from the $\sim$0.5~keV gas is detectable above the background below 
$\sim$1~keV.\label{fig_spectra}}
\end{figure*}
We extracted spectra in a series of concentric, contiguous annuli, placed at the X-ray centroid
(which was computed iteratively and verified by visual inspection). 
The widths of the annuli were chosen so as to contain approximately the same number of 
background-subtracted photons, and ensure there were sufficient photons to perform useful
spectral-fitting. The resulting annuli all had widths larger than $\sim$8\arcsec, which
is sufficient to prevent the instrumental spatial resolution from producing strong
mixing between the spectra in adjacent annuli.  The data in the vicinity of any
detected point source were excluded, as were the data from the vicinity
of chip gaps, where the instrumental response may be uncertain. We extracted
products from all the active chips (excluding the S4 CCD, which suffers from
considerable noise), making appropriate count-weighted spectral response
matrices for each annulus
with the standard \ciao\ tasks {\tt mkwarf} and {\tt mkacisrmf}. We extracted identical
products individually for each dataset, 
taking into account the astrometric offsets between them (determined from
the image registration discussed above). The spectra were
added  using the standard \heasoft\ task {\tt mathpha},  
and the response matrices were averaged using the \heasoft\ tasks
{\tt addrmf} and {\tt addarf}, with weights based on the number of photons in
each spectrum. Representative spectra, without background subtraction, are 
shown in Fig~\ref{fig_spectra}, along with the best-fitting source and background models.

\begin{figure*}
\centering
\includegraphics[width=7in]{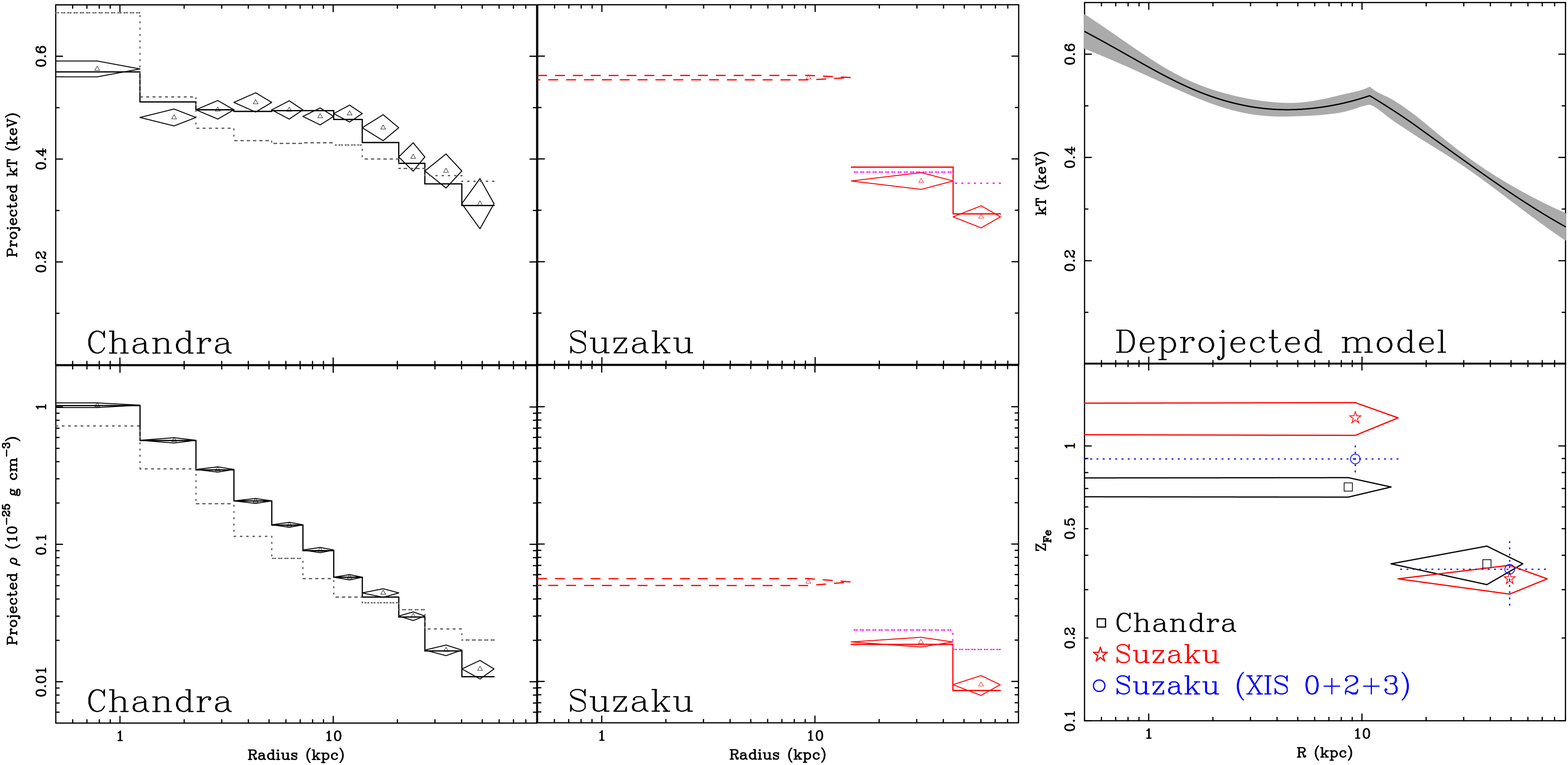}
\caption{Radial projected density, temperature and abundance profiles for \src, obtained
with both \chandra\ and \suzaku. (For the definition of projected density, see text).
Overlaid are the best hydrostatic models for each dataset, which match
the data very well. The dotted lines are the best models if 
dark matter is omitted, for which the fit is very poor. 
The temperature and density data in the central bin 
of the \suzaku\ profile were excluded from the fit as the effects of the telescope's limited
spatial resolution are most problematical at this scale, but are shown here as dashed 
diamonds. The central \suzaku\ abundance data-point appears sensitive to systematic
uncertainties (\S~\ref{sect_syserr_instrument}); we show the results for a simultaneous
fit of all the XIS units, and for a combination only of units 0, 2 and 3. 
The displayed error-bars do not take into account any covariance between data-points, which is 
accounted for in our analysis.
In the upper right panel, we show the best-fitting, deprojected temperature profile, and the 
corresponding 1-$\sigma$ confidence range (shaded region).} \label{fig_profiles}
\end{figure*}
Spectral-fitting was carried out in the energy-band 0.5--7.0~keV, using \xspec\ vers.~12.5.1n.
We have shown previously that fits to Poisson distributed data which minimize $\chi^2$ can yield 
significantly biased results, even if we rebin the data to more than the canonical $\sim$20 counts 
per bin \citep{humphrey09b}.
In contrast, we found that the C-statistic of \citet{cash79a} typically
gives relatively unbiased results.
We therefore performed the fits by minimizing C, but nonetheless 
took care to verify there is no significant bias, using the Monte Carlo procedure outlined in
\citet{humphrey09b}. Although not strictly necessary for a fit using the C-statistic, we rebinned
the data to ensure a minimum of 20 photons per bin. This aids convergence by emphasizing 
differences between the model and data. 
The data in all annuli were fitted simultaneously, to allow us to constrain the source and background
components at the same time.

We modelled the source emission in each annulus as coming from a single APEC plasma model 
with variable abundances, which was modified by Galactic foreground absorption \citep{dickey90}. 
Unlike several of our recent studies of galaxies \citep{humphrey06a,humphrey08a,humphrey09d,humphrey10a} 
we did {\em not} deproject the data, but instead fitted directly the projected spectra, similar
to \citet{gastaldello07a}. We adopted this procedure since we were interested in accurately tracing the 
gas density to the largest available radii; deprojection requires the accurate subtraction of the 
emission from beyond the outermost radius, which is challenging given the very flat surface brightness
profile of this galaxy. In our previous studies, our approximate method for achieving this introduced 
significant uncertainties into the outermost density data-point, which we have found to be prohibitive
here. Since the temperature profile is relatively isothermal, the 
emission-weighted, projected gas density and temperature
can be computed fairly accurately with the ``response weighting'' algorithm outlined in Appendix
B of \citeauthor{gastaldello07a}. We 
discuss the impact on our results of fitting the deprojected data in \S~\ref{sect_syserr_3d}.

To model the source spectrum, we adopted a slightly modified version of the \xspec\ {\em vapec}
model, for which the abundance {\em ratios} of each species with respect to Fe are directly
constrained. We allowed the abundance of Fe (\zfe), and the abundance ratios of O, Ne, Mg, Si and Ni
with respect to Fe to vary. The other abundances (He) or abundance ratios with respect to Fe (for
the other species) were fixed at their Solar values \citep{asplund04a}. 
To improve the constraints, \zfe\ was tied between multiple annuli, where required, and the 
abundance ratios were tied between all annuli. The best-fitting
abundances (Fig~\ref{fig_profiles}), are consistent (in the central part of the system) 
with previous \chandra\ results from the
shallow (17~ks) observation \citep{humphrey05a}, and are competitive with those obtained from
the very deep \suzaku\ observation (\S~\ref{sect_suzaku}; \citealt{tawara08a}), 
as well as the \xmm\ and \chandra\ measurements of 
the gas in much more massive (and X-ray bright) objects
\citep[\eg][]{humphrey05a,kim04a,buote03b}. Intriguingly, we note evidence of a significant
abundance gradient in this system (discussed in more detail in \S~\ref{sect_abundance_gradient}).

To account
for emission from undetected point sources, we included an additional 7.3~keV bremsstrahlung 
component. Since the number of X-ray point sources is approximately proportional to the 
stellar light \citep[\eg][]{humphrey08b}, the relative normalization of this component between each
annulus was fixed to match the relative K-band luminosity in the matching regions, which we 
measured from the \twomass\ image. To determine the total X-ray binary luminosity, we added the 
total luminosity of this component, summed over all annuli, to the total \lx\ of the detected
sources within the \dtwentyfive\ ellipse \citep{devaucouleurs91}\footnote{To obtain \lx\ for the 
detected sources, we fitted an absorbed bremsstrahlung model to their composite spectrum.}.
The resulting \lx,  $(2.95\pm 0.12)\times 10^{40}$\ergps, is
in excellent agreement with that inferred from extrapolating the point source luminosity function
\citep[$(2.8\pm 0.8)\times 10^{40}$\ergps:][]{humphrey08b}.

To account for the background, we included additional spectral components in our fits, 
in a variant of the modelling procedure outlined in \citetalias{humphrey06a}, 
which were fitted simultaneously to each spectrum, along with the source. To account for 
the sky background, we included two (unabsorbed) APEC components (kT=0.07~keV and 0.2~keV) and an
(absorbed) powerlaw component ($\Gamma=1.41$). The normalization of each component within each annulus
was assumed to scale with the extraction area, but the total normalizations were fitted freely. 
(We discuss the likely impact of the ``Solar wind charge exchange'' component in \S~\ref{sect_syserr_swcx}).
To account for the instrumental background, we included a number of Gaussian lines and a broken powerlaw model,
which were not folded through the ARF. We included separate instrumental components for the front- and 
back-illuminated chips, and assumed that the normalization of each component scaled with the area of the 
extraction annulus which overlapped the appropriate chips. The normalization of each component, and the shape
of the instrumental components, were allowed to fit freely.
We included two Gaussian lines (at 1.77 and $\sim$2.2~keV), the 
intrinsic widths of which were fixed to zero. The energies of the $\sim$2.2~keV lines were allowed to 
fit freely, as were as the normalizations of all the components.

The best-fitting models are shown in Fig~\ref{fig_spectra} for a representative 
selection of spectra. To verify the fit had not become trapped in a local minimum,
we explored the local parameter space by stepping individual parameters over
a range centred around the best-fitting value 
\citep[analogous to computing error-bars with the algorithm of][]{cash76}.
Error-bars were computed via the Monte Carlo technique outlined in \citet{humphrey06a}, 
and we carried out 150 error simulations.
In our previous work these simulations were used to generate
the standard error on each measured temperature and density data-point, which were assumed to be
uncorrelated. In a low-temperature (\ltsim 1~keV) system, however, there is a strong degeneracy between 
the abundance and the gas density. Tying the abundance between multiple annuli, as done in
the present work, will consequently 
result in correlated density errors. Therefore, rather than computing only
the error on each data-point from the error simulations, we also computed the covariance between each 
density data-point, which we fold into our mass modelling (see \S~\ref{sect_mass}). (In 
\S~\ref{sect_syserr_covariance}, we show that neglecting this effect, or adopting a full treatment 
that incorporates covariance between all the temperature and density
data-points, results in slightly different error-bars and a slight bias, but does not strongly influence
our conclusions.)
In  Fig~\ref{fig_profiles} we show the derived gas temperature, abundance and projected density profiles
and the 1-$\sigma$ error-bars on each data-point (assuming the data-points are uncorrelated).
The ``projected density'' is the mean gas density if all the emission measured in the annulus 
originates from a region defined by the intersection of a cylindrical
shell and a spherical shell, both of which have the inner and outer radii as the 
annulus. Defining it in this way is convenient as it has the units of density, but its precise 
definition is unimportant, so long as  care is taken to compute the models appropriately.

As is clear from Fig~\ref{fig_spectra}, the instrumental component dominates the 
background.
This reflects both the strong non X-ray background of the telescope, but also
the ability of \chandra\ to mitigate the cosmic component by resolving a significant fraction of it
into individual point sources, which were excluded from subsequent analysis.
Nevertheless, below $\sim$1~keV, the emission from the $\sim$0.5~keV gas in \src\ dominates 
at least out to $\sim$60~kpc, allowing the gas temperature and density to be constrained. 
Emission from unresolved point-sources associated with \src\ is important only in the innermost few bins,
and (as is clear from Fig~\ref{fig_spectra}) it can be clearly disentangled from the hot
gas component, as it has a very different spectral shape. We note that we have not 
explicitly included a component to account for the combined emission from cataclysmic variables
and hot stars \citep{revnivtsev08a}. This component, however, is expected to be an order
of magnitude fainter than the X-ray binary component, and so this omission will not 
affect our conclusions \citep{humphrey09d}.

\subsection{Suzaku} \label{sect_suzaku}
The observation of NGC\thin 720 with \suzaku\ was made early in the life of the satellite,
while all four XIS units were operating. Data-reduction was performed using the \heasoft\ 6.8
software suite, in conjunction with the XIS calibration database (\caldb) version 20090925.
To ensure up-to-date calibration, the unscreened data were re-pipelined with the {\tt aepipeline} task
and analysed following 
the standard data-reduction guidelines\footnote{{http://heasarc.gsfc.nasa.gov/docs/suzaku/analysis/abc/}}. 
Since the data for each instrument were divided into differently telemetered events file
formats, we converted the ``$5\times5$'' formatted data into  ``$3\times3$'' format,
and merged them with the ``$3\times3$'' events files.
The lightcurve of each instrument
was examined for periods of anomalously high background, but no significant amount of data
was found to be contaminated in this way. In Fig~\ref{fig_images}, we show the 0.5--7.0~keV image
for the XIS0 image, excluding data in the vicinity of the calibration sources. A number of 
off-axis point sources, most likely background AGNs, were identified by  visual inspection.
To minimize their contamination, we excluded data in a 1.5\arcmin\ {\em radius}
circular region centred on each of these sources. Given the large telescope point-spread 
function, such a cut will only eliminate $\sim$70\%\ of the photons from these sources.
However, we have chosen not to use a larger exclude region, since it will
cut out a large portion of the field of view while only slightly reducing the contamination.

Spectra were extracted in three concentric annuli (0--2\arcmin, 2--6\arcmin\ and 6--10\arcmin), 
centred at the nominal position of NGC\thin 720 in the field of view of each instrument.
Data in the vicinity of the calibration sources and the identified point sources were excluded.
For each spectrum, we generated an associated redistribution matrix file (RMF) using the {\tt xisrmfgen}
tool and an estimate of the instrumental background with the {\tt xisnxbgen}  task.
Since the 
point-spread function of the telescope is very large ($\sim$2\arcmin\ half power diameter), even with
such large apertures it is necessary to account for spectral mixing between each annulus.
We did this by 
generating multiple ancillary response files (ARFs) for each spectrum with the 
{\tt xissimarfgen} tool \citep{ishisaki07a}, which models the telescope's optics through ray-tracing.
Formally, we wish to model the true spectrum of the hot gas, 
$T^{gas}(\alpha,\delta,E)$, which is a function of right ascension ($\alpha$), declination ($\delta$) and 
energy ($E$). {To make the problem tractable, it is usual practice to adopt an approximation of 
the form $T^{gas}(\alpha,\delta,E)\simeq \sum_i a^{gas}_i(\alpha,\delta) t^{gas}_i(E) \xi^{gas} (\alpha,\delta)$, where $a^{gas}_i(\alpha,\delta)$ 
simply delineates discrete, non-overlapping regions of the sky (\ie\ $a^{gas}_i(\alpha,\delta)=1$ if 
($\alpha,\delta$) is within region $i$, or 0 otherwise), and we have assumed that, within region
$i$, the spectrum can be written $t^{gas}_i(E) \xi^{gas} (\alpha,\delta)$, where $t^{gas}_i(E)$ is 
the normalized spectrum, and $\xi^{gas}$ is the normalization 
as a function of position on the sky. This is analogous to expanding $T^{gas}(\alpha,\delta,E)$
on a set of basis functions that are separable in E and ($\alpha,\delta$). For our purposes, each $i$
refers to an annular region on the sky (with outer radii 2\arcmin, 6\arcmin\ and 10\arcmin), to 
correspond to the extraction regions. We approximated $\xi^{gas}$ as a $\beta$-model (with parameters
chosen to fit the \chandra\ image), correctly normalized to account for the gas emissivity.}
After passing through the telescope optics, a photon will be detected by the CCDs at some
detector coordinate (x,y) and pulse height (h; corresponding to the energy). The 
pulse height spectrum $P(x,y,h)$ can be written as \citep[\eg][]{davis01a}:
\begin{eqnarray}
P(x,y,h) =  \int d\alpha\ d\delta\ dE K(x,y,h,\alpha,\delta,E) T(\alpha,\delta,E) \nonumber
\end{eqnarray}
where $K$ encapsulates the effects of the telescope optics and the instrumental response. 
Now, we define $D_j(h)$ as the pulse height spectrum accumulated in an extraction region, $j$,
such that $D_j(h) = \int dx dy b_j(x,y) P(x,y,h)$. (Typically we would choose $i$ and $j$ to correspond
to the same set of regions on the sky, but that is not essential). 
Here $b_j(x,y)$ is 1 if $(x,y)$ lies within
region $j$, or 0 otherwise. Thus, we can write:
\begin{eqnarray}
D_j & = &   \sum_i \int dE\ t^{gas}_i  \left[ \int dx\ dy\ d\alpha\ d\delta\ a^{gas}_i b_j  \xi^{gas} K \right] \nonumber
\end{eqnarray}
Now, we assume that the term in square brackets can be replaced by a product of two
new terms, $R_j(E,h)$ and $A^{gas}_{ij}(E)$. With this definition, these new terms correspond,
respectively, to the redistribution matrix and ancillary response matrix that can be 
created with the tasks {\tt xisrmfgen} and {\tt xissimarfgen}. Hence, 
\begin{eqnarray}
D_j (h) = \sum_i \int dE\ t^{gas}_i(E) R_{j}(E,h) A^{gas}_{ij}(E) \label{eqn_suzaku}
\end{eqnarray}
Spectral fitting involves adopting a parameterized form for the spectrum $t_i(E)$ and comparing the 
corresponding model $D_j(h)$ to the observed pulse-height spectrum. Unlike \chandra, for which the
mixing between annuli was negligible (so that $A^{gas}_{ij}(E)=0$ if $i\ne j$), it was necessary to
to include all the ARFs $A^{gas}_{ij}$ in the \suzaku\ fitting, which can be done with \xspec. The spectra
from all extraction regions $j$ were fitted simultaneously, to allow all three $t_i(E)$ models to be constrained.
{We note that the \xspec\ {\em suzpsf} model can be used to perform a more approximate correction for
the mixing between annuli. However, our approach more formally accounts for the spatial distribution
of the source photons, and is more practial for our present purposes.}
The spectra from all instruments were fitted simultaneously, allowing
additional multiplicative constants in the fit to enable the relative normalization of the 
model for each instrument to vary with respect to the XIS 0.

It is necessary to add additional components to account for unresolved LMXBs and for the background
components. The ARF terms ($A_{ij}(E)$) depend not only on the extraction region
geometry but also the spatial distribution of the source photons, 
which clearly differs between the gas, LMXB and background components. Therefore, it is necessary 
to generate additional ARFs for each component. For each of these new components, we assume that the 
spectral shape does not vary with $\alpha$ and $\delta$, and so we can write
$T^{lmxb}(\alpha,\delta,E) = \xi^{lmxb}(\alpha,\delta) t^{lmxb}(E)$, and similar terms for the background.
We adopt a 7.3~keV bremsstrahlung model for $t^{lmxb}(E)$, and allow its normalization to be 
fitted freely.
Since LMXBs should trace the stellar light, we used the K-band \twomass\ image (appropriately normalized) 
to define the term $\xi^{lmxb}$.
Folding all of the components into the fitting, eqn~\ref{eqn_suzaku} is modified to:
\begin{eqnarray}
D_j(h) & = & \sum_i \int dE\ t^{gas}_i(E) R_{j}(E,h) A^{gas}_{ij}(E) \nonumber \\
       &   & + \int dE\ t^{lmxb}(E) R_j(E,h) A^{lmxb}_j(E) \nonumber \\
       &   & + \int dE\ t^{sky}(E) R_j(E,h) A^{sky}_j(E) \nonumber \\
       &   & + \int dE\ t^{inst}(E) \overline R_j(E,h)   \label{eqn_suzaku_full}
\end{eqnarray}
Here we consider separately the sky and instrumental (``inst'') background contributions. For the former,
we adopted two APEC plasma models plus a powerlaw term, as described in \S~\ref{sect_chandra_spectrum},
and assume that its surface brightness does not vary over the field of view
when computing $A^{sky}_j(E)$. For the latter term, we adopted two, complementary, approaches. 
One is to subtract off the 
spectrum generated for each annulus by {\tt xisnxbgen}, in which case the last term
in Eqn~\ref{eqn_suzaku_full} will be omitted. 
The other method, which we adopted by default as it 
provides greater generality in our fits, is to model the instrumental background, similarly to the treatment of this
 component with \chandra. We found that a single broken power law, plus Gaussian lines
(at $\sim$1.5, 1.7, 2.2, 5.9, 6.5 and 7.5~keV), was sufficient 
for our purposes here, provided the model was not multiplied by the effective area curve. 
Unfortunately, the quantum efficiency is
folded into the RMF created by {\tt xisrmfgen}, and so we first had to normalize the RMF through which this
model was folded (the ``renormalized'' response matrix being indicated by $\overline R_j$ in eqn~\ref{eqn_suzaku_full}). 
Still, we found that the precise shape of the instrumental term is not important since 
it is a relatively minor component in the spectral model (except in the vicinity 
of the calibration lines).
We did not obtain significantly different results if we used the modelling method or the 
direct subtraction method for the instrumental contribution (\S~\ref{sect_syserr_background}).

We found that our model was able to capture the shape of the spectra well. The best-fitting 
normalization of the LMXB component corresponds to an LMXB luminosity of 
$(3.08\pm0.07)\times 10^{40}$\ergps, in 
excellent agreement with the \chandra\ data, and with the extrapolation of the detected LMXB luminosity function.
Representative spectra are shown in Fig~\ref{fig_spectra}, showing the various model components. 
It is immediately clear that the spectral mixing between annuli is significant. For example, $\sim$20\%\ 
of the photons from the central bin are scattered into the second annulus. {\em Failure to take account
of this effect actually results in a significantly different temperature profile, which is almost isothermal at 
$\sim$0.5~keV, and very inconsistent with the \chandra\ data.} The background is dominated by the 
sky component, which is more significant than in \chandra\ since very few of the point sources that 
dominate it above $\sim$1~keV can be resolved and excluded. 

The hot gas temperature, abundance and projected density profiles are shown in Fig~\ref{fig_profiles}. 
In the 
two outer annuli, the data are in good agreement with the results from \chandra. 
In the innermost annulus, however, we found some discrepancies. Specifically, the abundance
is significantly higher than observed with \chandra, while the temperature is slightly higher than an
emission-weighted average of the \chandra\ temperature profile in this region.
The \suzaku\ results in the central bin are, however, sensitive to systematic uncertainties
in our modelling procedure (see \S~\ref{sect_syserr}). In particular, if we omitted data from the 
XIS1 unit from our fits, we obtained \zfe\ in much better agreement with \chandra\ 
(Fig~\ref{fig_profiles}). While this may point to problems with the calibration of the XIS1 unit,
it may also be related to errors in the correction we applied to account for spectral
mixing between the different annuli. If we omit this correction, for example, the discrepancy
with \chandra\ is significantly reduced (\zfe$\sim$0.9). 
{The real temperature profile is clearly more complex than we assume when 
performing this correction, which may be partially responsible, as there 
may be slight errors in the ray-tracing
algorithm. Furthermore, the single-temperature model fitted to the large \suzaku\ regions, within
which the \chandra\ data reveal temperature gradients, would not be expected to yield perfectly
consistent results with the \chandra\ data. }
In any case, we omit the data from the central \suzaku\ annulus in our subsequent modelling,
since the \chandra\ data are expected to be more reliable on these scales.

\section{Mass modelling} \label{sect_mass}
\begin{figure*}
\centering
\includegraphics[width=7in]{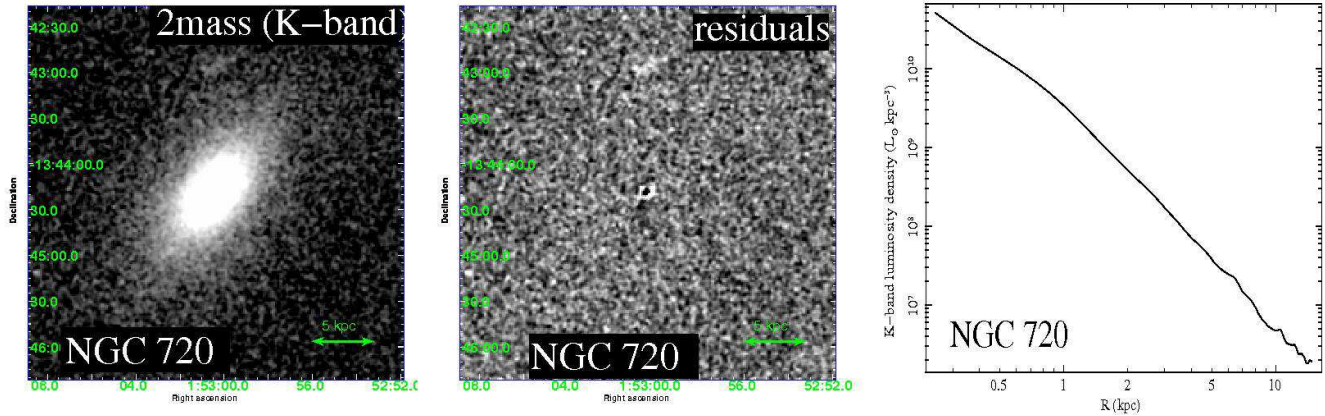}
\caption{{\em Left:} K-band \twomass\ image of \src. 
{\em Centre:} residual image, having subtracted our projected
model for the three-dimensional light distribution. Overall, this model provides a good fit to the stellar
light distribution. {A small, coherent residual feature is visible towards the top of the image, which may
not be associated with the galaxy. If it is a part of \src, its luminosity is too faint to affect our
conclusions.} {\em Right:} the spherically averaged mass density as a function
of 3-dimensional radius from our model.  \label{fig_optical_images}}
\end{figure*}
We translated the density and temperature profiles into mass constraints using the entropy-based
``forward-fitting'' technique outlined in \citetalias{humphrey08a} and 
\citet{humphrey09d}. Briefly, given parametrized profiles of 
``entropy'' (S=kT$n_e^{-2/3}$, where $n_e$ is the electron density) 
and gravitating mass (excluding the gas mass, which is 
computed self-consistently), plus the gas density at some canonical  radius
(for which we used 10~kpc),
the three-dimensional temperature and density profiles can be calculated, under the 
hydrostatic equilibrium approximation and fitted to the data. 
This model assumes spherical symmetry, which is a standard
assumption in X-ray hydrostatic modelling, even when the X-ray isophotes are not perfectly 
round, since deviations from sphericity are expected to contribute only a relatively small
error into the recovered mass profile \citep[\eg][]{piffaretti03a,gavazzi05a}, and the inferred 
baryon fraction \citep{buote10a}.

{For the gravitating mass model (in addition to the gas mass), we explored two possible 
models. Firstly, we considered a model without dark matter, so that the 
gravitating mass comprised only a stellar component, a supermassive black hole
\citep[with mass fixed at $3\times 10^8$\msun, which is that estimated from the \mbh-\sigmac\ relation of][given the central velocity dispersion value listed in the HyperLEDA database]{tremaine02a},
and the mass of the gas. We assumed the stellar mass component followed the optical light, but allowed the mass-to-light
(M/L) ratio to be fitted freely. In the second model, we also included an NFW 
\citep{navarro97} dark matter
halo, the virial mass and concentration (defined as the ratio of the virial radius
to the characteristic scale of the NFW model) of which were allowed to fit freely.}
 To model the stellar light, we
deprojected the K-band \twomass\ image using 
the method outlined in \citet{humphrey09d}, which is based on the algorithm of
\citet{binney90a}.  This entailed first fitting an arbitrary, 
analytical model to the \twomass\ image, which was deprojected analytically onto an 
arbitrary axisymmetric grid, for a given inclination angle. Since NGC\thin 720
exhibits a very elliptical optical morphology, it is improbable that it is being
observed at a low (face-on) inclination angle, and so we initially assumed 
a 90$^\circ$ inclination (but investigate the effect of using lower inclinations in
\S~\ref{sect_syserr_stars}).  A series of  \citet{lucy74a} relaxation iterations
gradually improved the deprojection.
In \citet{humphrey09d}, we demonstrated that 
this algorithm gives results consistent with the alternative deprojection technique of
\citet{gebhardt96a}. Since the innermost region within which we extracted the \chandra\
data  has a radius of $\sim$10\arcsec, it was not necessary to deconvolve the 
seeing from the \twomass\ image prior to deprojection.
We show in Fig~\ref{fig_optical_images} the K-band 
\twomass\ image, the residuals from the projection of our axisymmetric model (demonstrating
the excellent fit), and the spherically averaged stellar light density profile. 
Although this method yields an axisymmetric three-dimensional light profile, following 
\citet{humphrey09d}, we subsequently averaged it spherically for use with our 
X-ray modelling \citep[see][]{buote10a}.

{To fit the three-dimensional entropy profile we assumed a model comprising a broken 
powerlaw, plus a constant; such a model is reasonably successful at reproducing the 
entropy profiles of galaxies and galaxy groups over a wide radial range 
\citep{humphrey09d,jetha07a,finoguenov07a,gastaldello07b,sun08a}.
The normalization of the powerlaw and constant components, the radius of the break
and the slopes above and below it were allowed to fit freely. We investigate the
impact of varying the slope outside the range of the data ($\sim$80~kpc)
between reasonable limits, which should bound all plausible extrapolated entropy
profiles, in \S~\ref{sect_syserr_entropy}.}

Following \citet{humphrey08a}, we solved the equation of hydrostatic equilibrium to determine
the gas properties as a function of radius, from 10~pc to a large radius outside the 
field of view. 
For the latter, we adopted the virial radius of the system defined by ignoring the
baryonic components (which is slightly smaller than the true virial radius of the 
system)\footnote{For the constant M/L ratio model, we found it more convenient to use
a less conservative, smaller radius. Specifically we used 75~kpc, and found the 
model agreed even less well if a larger scale was adopted.}. 
For any arbitrary set of mass and entropy model parameters, it is not always possible to 
find a physical solution to this equation over the full radial range. Such 
models were therefore rejected as unphysical during parameter space exploration. 
In order to compare to the observed {\em projected} density and temperature profiles, we 
projected the three-dimensional gas density and temperature, using a procedure
similar to that described in \citet{gastaldello07a}\footnote{In computing the plasma emissivity 
term, we approximated the true three dimensional abundance profile with the projected abundance profile
(Fig~\ref{fig_profiles}); since the  profile is relatively flat, this should be sufficient
for our present purposes.}.
This involves computing an emission-weighted projected mean temperature and density in each 
radial bin, that can be compared directly to the data\footnote{We note that, for systems with
a strong temperature gradient and  
kT \gtsim 3~keV the emission-weighted temperature
may be biased low \citep{mazzotta04a}. For the temperature of NGC\thin 720 the effect is expected to be
be weak \citep{buote00a}, which we confirm given the lack of any significant bias we find when fitting the deprojected 
data, rather than the projected data (\S~\ref{sect_syserr_3d}).}.

To compare the model to the data, we used the $\chi^2$ statistic, fitting simultaneously the 
\chandra\ and \suzaku\ temperature and density profiles, with the central bin of the 
\suzaku\ data excluded from our fits. Correlations between the density errors were simply 
implemented by adopting a form for the $\chi^2$ statistic which incorporates the covariance
matrix \citep[\eg][]{gould03a}\footnote{By default we only consider correlations between
the density data, but we investigate introducing a more complete covariance computation in 
\S~\ref{sect_syserr_covariance} and find that the results are not significantly affected.}. The probability density function (of the data, given the model)
is modified in a similar manner when folding in the non-diagonal elements of the covariance matrix. 
Parameter space was explored with a Bayesian Monte Carlo method. Specifically, we used 
version 2.7 of the MultiNest code\footnote{{http://www.mrao.cam.ac.uk/software/multinest/}} \citep{feroz08a,feroz09a}.
Since the choice of priors is nontrivial, we followed convention in cycling through a 
selection of different priors and assessing their impact on our results 
(we discuss this in detail in \S~\ref{sect_syserr_priors}). Initially, however, we adopted flat priors on
the logarithm of the dark matter mass (over the range $10^{12}< M < 10^{16}$\msun), 
the logarithm of the dark matter halo concentration, $c_{DM}$ (over the range $1<c_{DM}<100$),
the logarithm of the gas density at the canonical radius, the stellar M/\lk\ ratio, and the various entropy 
parameters. For a more detailed description of the modelling
procedure, see \citet{humphrey08a,humphrey09d}. 

\section{Results}
\begin{figure}
\includegraphics[width=3.4in]{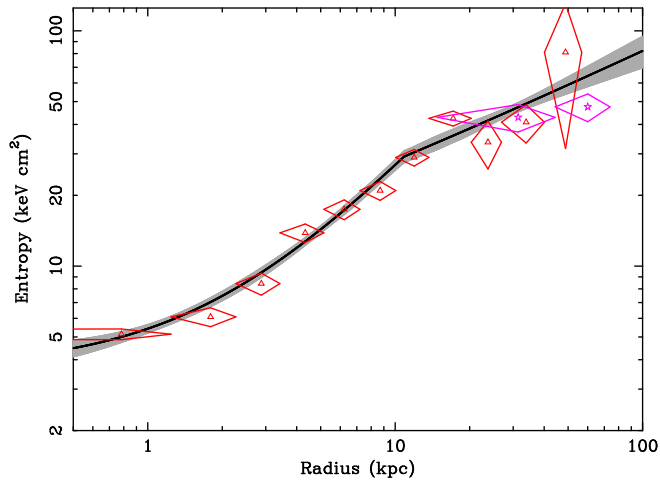}
\caption{Best-fitting entropy profile from fitting the 
projected data (solid line). The grey shaded region indicates the
approximate 1-$\sigma$ scatter. The sharpness of the break at $\sim$14~kpc
is a consequence of the simple, broken powerlaw parameterization we adopted
for the entropy profile. Overlaid are data-points obtained
directly from the deprojected \chandra\ (triangles) and 
\suzaku\ (stars) data (\S~\ref{sect_syserr_3d}),
which agree well with the projected fit.
The deprojected data points have much larger errors as deprojection
tends to amplify noise, and the model fit uses both \chandra\
and \suzaku\ data.\label{fig_entropy}}
\end{figure}
\begin{deluxetable}{lrrl}
\tablecaption{Entropy and pressure fit results\label{table_entropy}}
\tablewidth{240pt}
\tablehead{
\colhead{Parameter} & \colhead{Marginalized value} & \colhead{Best-fit} & \colhead{units}
}
\startdata
$\log_{10} \rho_{g,10}$ & -26.27$\pm0.02$ & (-26.28) & [g cm$^{-3}$]\\
$S_c$ & 3.9$\pm$0.6 & (3.7) & keV cm$^{2}$ \\
$S_0$ & 23$\pm$2 & (23) & keV cm$^{2}$ \\
$\beta_1$ & 1.14$\pm0.19$ & (1.13)& \\
$R_{brk}$ & 11.6$^{+3.0}_{-2.4}$ & (10.9)& kpc \\
$\beta_2$ & 0.47$\pm 0.13$ & (0.51) & 
\enddata
\tablecomments{Marginalized value and 1-$\sigma$ errors 
on the  deprojected gas density at 10~kpc ($\rho_{g,10}$),
and the entropy parameters. The fitted entropy model within 100~kpc has the 
form $S(R)=S_c+S_0 f(R)$, where $f(R)=R_{10}^{\beta_1}$, if $R<R_{brk}$,
or else $f(R)=(R_{brk}/10 kpc)^{\beta1-\beta2} R_{10}^{\beta_2}$, and
$R_{10}=R/10\ kpc$. Since the best-fitting set of parameters need not be
identical to the marginalized values, we also list the best-fitting value for 
each parameter.}
\end{deluxetable}
\begin{figure}
\includegraphics[width=3.4in]{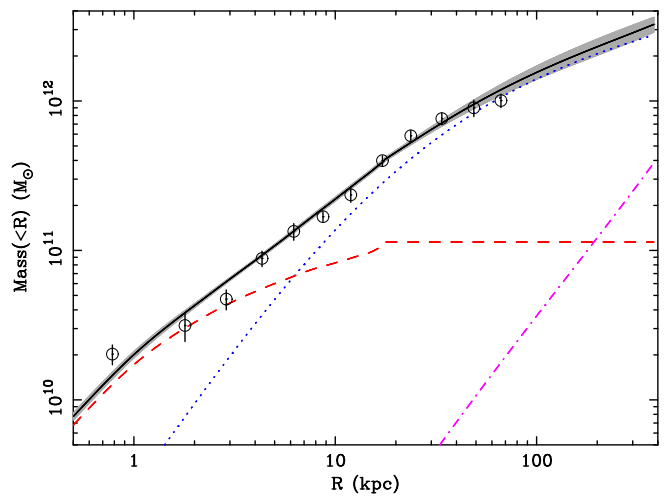}
\caption{Radial mass profile of \src. 
The solid (black) line indicates the total enclosed
mass, the dashed (red) line indicates the stellar mass, the dotted
(blue) line is the dark matter contribution and the dash-dot (magenta) line
is the gas mass contribution. The grey shaded regions indicate the
1-$\sigma$ error on the total mass distribution. Overlaid are a
set of data-points derived from the more ``traditional'' mass
analysis of this system described in \citet{humphrey10a}, which generally agrees very well
with the fitted model (we stress the model is {\em not} fitted to
these data-points, but is derived separately). Note that the kink in the 
stellar mass model at $\sim$20 kpc is an artefact of the stellar light 
deprojection technique, but does not affect our results (\S~\ref{sect_syserr_stars}).
\label{fig_mass_profile}}
\end{figure}
\begin{figure}
\includegraphics[width=3.4in]{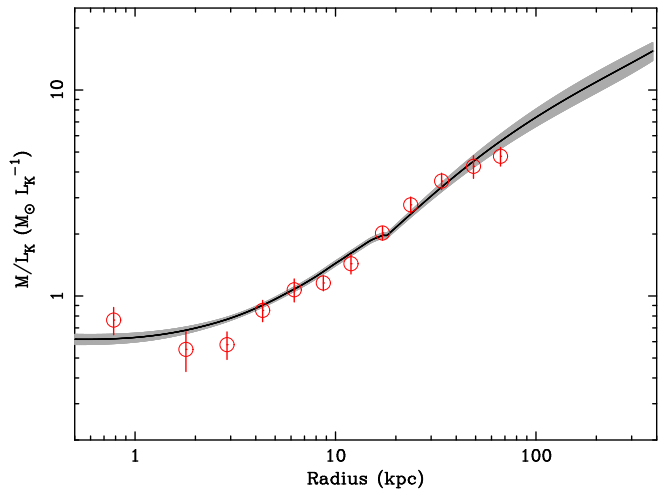}
\caption{Radial total mass-to-K-band light profile of \src. The shaded region
indicates the 1-$\sigma$ scatter about the best-fitting relation. The steep rise 
with radius outside $\sim$\reff\ indicates the presence of a massive DM halo. The small
kink in the profile around $\sim$20~kpc is an artefact of the stellar light deprojection
technique, and is unrelated to the break in the entropy profile at $\sim$10~kpc. Overlaid 
are a series of data-points derived from the more traditional mass analysis
shown in Fig~\ref{fig_mass_profile}. We stress that the data-points are derived 
independently and the model is {\em not} fitted to these data.
\label{fig_mass_to_light}}
\end{figure}

\subsection{No dark matter model}
{We first consider the case without a dark matter halo component. Although our 
previous work (\eg\ \citetalias{humphrey06a}) has found that the M/L ratio
rises steeply with radius (implying that dark matter is needed at high significance),
given the high-quality data in \src, it is of particular 
interest to quantify how poorly the constant M/L ratio model fits these data.
We find that the best-fitting model without dark matter is a very poor 
description of the temperature and density profiles (Fig~\ref{fig_profiles}; 
$\chi^2$/dof=413/19, with a mean absolute fractional deviation between
the data and models of $\sim$29\%). As found by \citet{buote02b}, based on
an analysis of the X-ray image in \src, the stellar mass component is too
centrally concentated to produce the observed gas pressure profile.}

\subsection{Dark matter halo} \label{sect_results_dm}
{As is clear in Fig~\ref{fig_profiles}, when a dark matter halo component 
was added, the hydrostatic model was able to capture
the overall shape of the projected density and temperature profiles very well
($\chi^2$/dof=12.7/17). The improvement to the fit when dark matter was included
was therefore highly significant; the ratio of the Bayesian ``evidence'' returned by the
fitting algorithm \citep[see][]{feroz08a} for the two cases is $10^{-85}$,
implying that dark matter is required\footnote{Although not
correct in this case \citep{protassov02}, for
reference, 
it is interesting to quote the significance of the dark matter detection in frequentist
terms by using the F-test; in this case, such an improvement in $\chi^2$ would occur
by random chance with a probability of 1.4$\times 10^{-13}$,
implying a $\sim$7.4-$\sigma$ detection of dark matter.} at $\sim$19.6-$\sigma$. }

\begin{deluxetable*}{llllllll}
\tablecaption{Mass results and error budget\label{table_mass}}
\tabletypesize{\scriptsize}
\tablehead{
\colhead{Test} & \colhead{M$_*$/\lk} & \colhead{log \mtwentyfive} & \colhead{log $c_{2500}$}& \colhead{log \mfive} & \colhead{log $c_{500}$} & \colhead{log \mvir} & \colhead{log \cvir} \\ 
& \colhead{\msun\lsun$^{-1}$}& \colhead{[\msun]} & & \colhead{[\msun]} & & \colhead{[\msun]}  
}
\startdata
Marginalized & $0.54^{+0.03}_{-0.05}$& $12.18^{+0.06}_{-0.05}$& $0.86 \pm 0.08$& $12.34^{+0.07}_{-0.05}$& $1.15 \pm 0.07$& $12.49^{+0.06}_{-0.04}$& $1.43 \pm 0.08$\\
 Best-fit& $(0.54)$& $(12.20)$& $(0.85)$& $(12.36)$& $(1.14)$& $(12.51)$& $(1.42)$\\\hline
\multicolumn{8}{c}{Systematic error estimates} \\
$\Delta$DM&  $+0.07$  $\left( ^{+0.03}_{-0.05}\right)$  &  $+0.12$  $\left( \pm {0.03}\right)$  &  $+0.46$  $\left( \pm {0.08}\right)$  &  $+0.30$  $\left( \pm {0.03}\right)$  &  $+0.52$  $\left( \pm {0.08}\right)$  &  $+0.48$  $\left( \pm {0.03}\right)$  &  $+0.58$  $\left( \pm {0.08}\right)$ \\
$\Delta$AC&  $-0.165$  $\left( \pm {0.04}\right)$  &  $+0.03$  $\left( \pm {0.06}\right)$  &  $-0.092$  $\left( \pm {0.09}\right)$  &  $+0.05$  $\left( \pm {0.07}\right)$  &  $-0.087$  $\left( \pm {0.09}\right)$  &  $+0.05$  $\left( \pm {0.06}\right)$  &  $-0.089$  $\left( \pm {0.09}\right)$ \\
$\Delta$Background&  $-0.077$  $\left( \pm {0.05}\right)$  &  $-0.051$  $\left( \pm {0.04}\right)$  &  $+0.11$  $\left( ^{+0.07}_{-0.08}\right)$  &  $-0.059$  $\left( \pm {0.05}\right)$  &  $+0.10$  $\left( \pm {0.07}\right)$  &  $-0.059$  $\left( ^{+0.05}_{-0.04}\right)$  &  $+0.10$  $\left( \pm {0.07}\right)$ \\
$\Delta$SWCX&  $-0.057$  $\left( \pm {0.05}\right)$  &  $-0.059$  $\left( ^{+0.05}_{-0.03}\right)$  &  $+0.10$  $\left( \pm {0.08}\right)$  &  $-0.055$  $\left( \pm {0.05}\right)$  &  $+0.10$  $\left( ^{+0.07}_{-0.08}\right)$  &  $-0.078$  $\left( ^{+0.05}_{-0.03}\right)$  &  $+0.10$  $\left( \pm {0.08}\right)$ \\
$\Delta$Entropy &  $-0.014$  $\left( \pm {0.04}\right)$  &  $+0.005$  $\left( \pm {0.06}\right)$  &  $\pm 0.01$  $\left( \pm {0.08}\right)$  &  $+0.008$  $\left( \pm {0.07}\right)$  &  $^{+0.01}_{-0.01}$  $\left( \pm {0.08}\right)$  &  $-0.033$  $\left( ^{+0.09}_{-0.06}\right)$  &  $^{+0.02}_{-0.02}$  $\left( \pm {0.09}\right)$ \\
$\Delta$3d &  $-0.008$  $\left( \pm {0.05}\right)$  &  $-0.016$  $\left( \pm {0.07}\right)$  &  $+0.009$  $\left( \pm {0.10}\right)$  &  $-0.031$  $\left( ^{+0.10}_{-0.06}\right)$  &  $+0.01$  $\left( \pm {0.10}\right)$  &  $-0.031$  $\left( ^{+0.08}_{-0.04}\right)$  &  $+0.02$  $\left( ^{+0.09}_{-0.12}\right)$ \\
$\Delta$Spectral&  $\pm 0.04$  $\left( \pm {0.06}\right)$  &  $^{+0.17}_{-0.03}$  $\left( ^{+0.08}_{-0.06}\right)$  &  $^{+0.05}_{-0.03}$  $\left( \pm {0.09}\right)$  &  $^{+0.16}_{-0.03}$  $\left( \pm {0.08}\right)$  &  $^{+0.05}_{-0.03}$  $\left( \pm {0.09}\right)$  &  $^{+0.15}_{-0.02}$  $\left( \pm {0.08}\right)$  &  $^{+0.05}_{-0.04}$  $\left( \pm {0.09}\right)$ \\
$\Delta$Weighting&  $-0.011$  $\left( ^{+0.05}_{-0.09}\right)$  &  $+0.01$  $\left( \pm {0.05}\right)$  &  $+0.06$  $\left( ^{+0.10}_{-0.12}\right)$  &  $+0.01$  $\left( \pm {0.06}\right)$  &  $+0.06$  $\left( ^{+0.09}_{-0.12}\right)$  &  $+0.006$  $\left( \pm {0.06}\right)$  &  $+0.06$  $\left( ^{+0.09}_{-0.12}\right)$ \\
$\Delta$Fit priors&  $^{+0.04}_{-0.02}$  $\left( \pm {0.05}\right)$  &  $^{+0.07}_{-0.05}$  $\left( \pm {0.06}\right)$  &  $^{+0.05}_{-0.12}$  $\left( \pm {0.09}\right)$  &  $^{+0.09}_{-0.05}$  $\left( \pm {0.07}\right)$  &  $^{+0.04}_{-0.12}$  $\left( \pm {0.08}\right)$  &  $^{+0.09}_{-0.05}$  $\left( \pm {0.06}\right)$  &  $^{+0.04}_{-0.12}$  $\left( \pm {0.08}\right)$ \\
$\Delta$Instrument&  $^{+0.01}_{-0.02}$  $\left( \pm {0.05}\right)$  &  $+0.10$  $\left( ^{+0.10}_{-0.07}\right)$  &  $-0.087$  $\left( ^{+0.08}_{-0.10}\right)$  &  $+0.11$  $\left( ^{+0.11}_{-0.09}\right)$  &  $-0.087$  $\left( ^{+0.07}_{-0.09}\right)$  &  $+0.11$  $\left( ^{+0.10}_{-0.07}\right)$  &  $-0.086$  $\left( \pm {0.09}\right)$ \\
$\Delta$Stars &  $^{+0.12}_{-0.03}$  $\left( \pm {0.33}\right)$  &  $^{+0.00}_{-0.01}$  $\left( ^{+0.08}_{-0.06}\right)$  &  $+0.01$  $\left( ^{+0.18}_{-0.26}\right)$  &  $^{+0.00}_{-0.02}$  $\left( ^{+0.12}_{-0.06}\right)$  &  $+0.01$  $\left( ^{+0.17}_{-0.25}\right)$  &  $-0.018$  $\left( ^{+0.11}_{-0.06}\right)$  &  $+0.02$  $\left( ^{+0.17}_{-0.25}\right)$ \\
$\Delta$Distance &  $\pm 0.04$  $\left( ^{+0.04}_{-0.05}\right)$  &  $^{+0.03}_{-0.01}$  $\left( \pm {0.06}\right)$  &  $^{+0.02}_{-0.04}$  $\left( \pm {0.08}\right)$  &  $^{+0.04}_{-0.01}$  $\left( \pm {0.06}\right)$  &  $^{+0.02}_{-0.04}$  $\left( \pm {0.08}\right)$  &  $\pm 0.03$  $\left( ^{+0.06}_{-0.05}\right)$  &  $^{+0.02}_{-0.04}$  $\left( \pm {0.08}\right)$ \\
$\Delta$Fit radius&  $+0.01$  $\left( ^{+0.05}_{-0.04}\right)$  &  $+0.09$  $\left( \pm {0.11}\right)$  &  $-0.072$  $\left( ^{+0.09}_{-0.12}\right)$  &  $+0.10$  $\left( \pm {0.12}\right)$  &  $-0.068$  $\left( ^{+0.08}_{-0.12}\right)$  &  $+0.09$  $\left( ^{+0.13}_{-0.10}\right)$  &  $-0.076$  $\left( ^{+0.09}_{-0.11}\right)$ \\
$\Delta$Covariance &  $-0.019$  $\left( \pm {0.06}\right)$  &  $^{+0.01}_{-0.01}$  $\left( ^{+0.06}_{-0.07}\right)$  &  $-0.015$  $\left( \pm {0.10}\right)$  &  $+0.02$  $\left( ^{+0.06}_{-0.08}\right)$  &  $-0.020$  $\left( ^{+0.11}_{-0.08}\right)$  &  $^{+0.01}_{-0.01}$  $\left( \pm {0.06}\right)$  &  $-0.022$  $\left( ^{+0.12}_{-0.08}\right)$
\enddata
\tablecomments{{Marginalized values and 1-$\sigma$ confidence regions for the stellar 
mass-to-light (M$_*$/\lk) ratio and the enclosed mass and concentration measured at various overdensities. 
Since the best-fitting parameters need not be identical to the marginalized
values, we also list the best-fitting values for each parameter (in parentheses). }
In addition to the statistical errors, we also show estimates of the error budget
from possible sources of systematic uncertainty.
We consider a range of different
systematic effects, which are described in detail in \S~\ref{sect_syserr}; 
specifically we evaluate the effect of the
choice of dark matter halo model ($\Delta$DM), adiabatic contraction ($\Delta$AC),
treatment of the background ($\Delta$Background) and the Solar wind charge exchange
X-ray component ($\Delta$SWCX), the extrapolation of the entropy model ($\Delta$Entropy), 
deprojection ($\Delta$3d), the employed
spectral model ($\Delta$Spectral), removing the emissivity correction 
($\Delta$Weighting), varying the priors on the model parameters ($\Delta$Fit priors)
the X-ray detectors used ($\Delta$Instrument),
the modelling of the stellar light ($\Delta$Stars), distance uncertainties
($\Delta$Distance), fit radius ($\Delta$Fit radius) and covariance between the 
temperature and density data-points ($\Delta$Covariance).  We list the change
in the marginalized value of each parameter for every test and, in parentheses,
the statistical uncertainty on the parameter determined from the test.
Note that the systematic error estimates should {\em not}
in general be added in quadrature with the statistical error.}
\end{deluxetable*}

\begin{deluxetable*}{llllrrr}
\tablecaption{Baryon fraction results and error budget\label{table_fb}}
\tabletypesize{\scriptsize}
\tablehead{
\colhead{Test} & \colhead{$f_{g,2500}$} & \colhead{$f_{g,500}$} & \colhead{$f_{g,vir}$}& \colhead{$f_{b,2500}$} & \colhead{$f_{b,500}$} & \colhead{$f_{b,vir}$}\\
}
\startdata
Marginalized & $0.027^{+0.003}_{-0.004}$& $0.06^{+0.009}_{-0.02}$& $0.12^{+0.05}_{-0.03}$& $0.10 \pm 0.01$& $0.11^{+0.02}_{-0.02}$& $0.16 \pm 0.04$\\
Best-fit& $(0.024)$& $(0.053)$& $(0.119)$& $(0.096)$& $(0.102)$& $(0.154)$\\\hline
\multicolumn{7}{c}{Systematic error estimates} \\
$\Delta$DM profile&  $-0.004$  $\left( \pm {0.003}\right)$  &  $-0.024$  $\left( \pm {0.01}\right)$  &  $-0.085$  $\left( ^{+0.01}_{-0.01}\right)$  &  $-0.013$  $\left( \pm {0.01}\right)$  &  $-0.044$  $\left( \pm {0.01}\right)$  &  $-0.114$  $\left( \pm {0.01}\right)$ \\
$\Delta$AC&  $-0.001$  $\left( \pm {0.004}\right)$  &  $-0.010$  $\left( ^{+0.01}_{-0.01}\right)$  &  $-0.019$  $\left( ^{+0.05}_{-0.03}\right)$  &  $-0.026$  $\left( \pm {0.01}\right)$  &  $-0.023$  $\left( \pm {0.02}\right)$  &  $-0.035$  $\left( \pm {0.04}\right)$ \\
$\Delta$Background&  $-0.003$  $\left( \pm {0.003}\right)$  &  $-0.001$  $\left( \pm {0.01}\right)$  &  $+0.03$  $\left( \pm {0.04}\right)$  &  $-0.005$  $\left( ^{+0.01}_{-0.01}\right)$  &  $\pm 0$  $\left( \pm {0.02}\right)$  &  $+0.02$  $\left( ^{+0.04}_{-0.03}\right)$ \\
$\Delta$SWCX&  $\pm 0$  $\left( \pm {0.002}\right)$  &  $+0.001$  $\left( ^{+0.01}_{-0.01}\right)$  &  $+0.03$  $\left( ^{+0.03}_{-0.04}\right)$  &  $+0.003$  $\left( \pm {0.01}\right)$  &  $+0.005$  $\left( \pm {0.01}\right)$  &  $+0.03$  $\left( ^{+0.03}_{-0.04}\right)$ \\
$\Delta$Entropy &  $\pm 0$  $\left( \pm {0.004}\right)$  &  $-0.018$  $\left( \pm {0.01}\right)$  &  $^{+0.02}_{-0.04}$  $\left( \pm {0.05}\right)$  &  $\pm 0$  $\left( \pm {0.01}\right)$  &  $^{+0.01}_{-0.01}$  $\left( \pm {0.02}\right)$  &  $^{+0.01}_{-0.05}$  $\left( \pm {0.05}\right)$ \\
$\Delta$3d &  $+0.005$  $\left( ^{+0.004}_{-0.005}\right)$  &  $+0.010$  $\left( \pm {0.02}\right)$  &  $+0.04$  $\left( ^{+0.05}_{-0.08}\right)$  &  $+0.01$  $\left( \pm {0.01}\right)$  &  $+0.01$  $\left( \pm {0.03}\right)$  &  $+0.03$  $\left( ^{+0.06}_{-0.08}\right)$ \\
$\Delta$Spectral&  $^{+0.001}_{-0.008}$  $\left( \pm {0.003}\right)$  &  $-0.026$  $\left( \pm {0.01}\right)$  &  $^{+0.02}_{-0.07}$  $\left( \pm {0.05}\right)$  &  $^{+0.00}_{-0.03}$  $\left( ^{+0.01}_{-0.01}\right)$  &  $^{+0.01}_{-0.04}$  $\left( \pm {0.02}\right)$  &  $^{+0.03}_{-0.09}$  $\left( \pm {0.05}\right)$ \\
$\Delta$Weighting&  $-0.001$  $\left( \pm {0.002}\right)$  &  $\pm 0$  $\left( \pm {0.01}\right)$  &  $+0.01$  $\left( \pm {0.05}\right)$  &  $-0.006$  $\left( \pm {0.01}\right)$  &  $-0.003$  $\left( \pm {0.01}\right)$  &  $+0.008$  $\left( \pm {0.05}\right)$ \\
$\Delta$Fit priors&  $^{+0.007}_{-0.004}$  $\left( \pm {0.004}\right)$  &  $^{+0.01}_{-0.02}$  $\left( \pm {0.02}\right)$  &  $^{+0.02}_{-0.04}$  $\left( ^{+0.06}_{-0.05}\right)$  &  $^{+0.01}_{-0.01}$  $\left( \pm {0.01}\right)$  &  $\pm 0.02$  $\left( \pm {0.02}\right)$  &  $^{+0.01}_{-0.06}$  $\left( \pm {0.06}\right)$ \\
$\Delta$Instrument&  $-0.008$  $\left( \pm {0.005}\right)$  &  $-0.025$  $\left( \pm {0.02}\right)$  &  $-0.067$  $\left( \pm {0.06}\right)$  &  $-0.021$  $\left( \pm {0.01}\right)$  &  $-0.032$  $\left( \pm {0.02}\right)$  &  $-0.081$  $\left( \pm {0.05}\right)$ \\
$\Delta$Stars &  $\pm 0$  $\left( \pm {0.004}\right)$  &  $-0.006$  $\left( ^{+0.02}_{-0.01}\right)$  &  $+0.008$  $\left( \pm {0.05}\right)$  &  $^{+0.02}_{-0.01}$  $\left( ^{+0.03}_{-0.05}\right)$  &  $-0.008$  $\left( ^{+0.03}_{-0.02}\right)$  &  $^{+0.01}_{-0.01}$  $\left( ^{+0.05}_{-0.06}\right)$ \\
$\Delta$Distance &  $-0.002$  $\left( ^{+0.004}_{-0.003}\right)$  &  $-0.007$  $\left( ^{+0.02}_{-0.01}\right)$  &  $+0.001$  $\left( ^{+0.05}_{-0.04}\right)$  &  $^{+0.00}_{-0.00}$  $\left( \pm {0.01}\right)$  &  $-0.006$  $\left( ^{+0.02}_{-0.02}\right)$  &  $-0.006$  $\left( \pm {0.05}\right)$ \\
$\Delta$Fit radius&  $-0.007$  $\left( ^{+0.007}_{-0.005}\right)$  &  $-0.019$  $\left( \pm {0.02}\right)$  &  $-0.064$  $\left( ^{+0.07}_{-0.04}\right)$  &  $-0.017$  $\left( \pm {0.02}\right)$  &  $-0.032$  $\left( ^{+0.03}_{-0.02}\right)$  &  $-0.076$  $\left( ^{+0.08}_{-0.04}\right)$ \\
$\Delta$Covariance &  $-0.002$  $\left( ^{+0.004}_{-0.003}\right)$  &  $-0.005$  $\left( \pm {0.01}\right)$  &  $+0.02$  $\left( ^{+0.04}_{-0.06}\right)$  &  $+0.002$  $\left( \pm {0.01}\right)$  &  $\pm 0.00$  $\left( \pm {0.02}\right)$  &  $+0.003$  $\left( \pm {0.05}\right)$
\enddata
\tablecomments{Marginalized values and 1-$\sigma$ confidence regions for the gas 
fraction ($f_{g,\Delta}$) and baryon fraction ($f_{b,\Delta}$) measured at various overdensities
($\Delta$). We also provide the best-fitting parameters in parentheses, and a breakdown 
of possible sources of systematic uncertainty, following Table~\ref{table_mass}. We find that
$f_b$ is reasonably robust to most sources of systematic uncertainty, especially within 
\rtwentyfive.}
\end{deluxetable*}
{In Fig~\ref{fig_entropy} we show the best-fitting entropy profile (and its 1-$\sigma$ confidence
range), which shows a flattening at large radius similar to two of the three galaxies
studied in \citet{humphrey09d}. As we will show later (\S~\ref{sect_entropy}), the entropy profile shape,
even when extrapolated to large radius, is consistent with trends seen in galaxy clusters
\citep{pratt10a}.
The derived radial distribution of the gravitating mass is shown in Fig~\ref{fig_mass_profile}
and the total mass-to-light ratio is shown in Fig~\ref{fig_mass_to_light}. In 
Fig~\ref{fig_mass_profile} we also show
the contribution to the mass from each separate component}. We find that the 
stars dominate the mass distribution within the central $\sim$5~kpc, 
but a significant dark matter halo is required to reproduce the data at larger radii,
consistent with our previous study of this system, using data from a much shallower 
observation
\citep{humphrey06a}.  The hot gas becomes the dominant baryonic component
outside $\sim$200~kpc ($\sim$\rfive). Overlaid are a series of mass data-points
derived from fitting the {\em deprojected} \chandra\ data using a more 
traditional (but more uncertain) 
mass-modelling method \citep[for more details see 
\S~\ref{sect_syserr_3d} and][]{humphrey09d}, which agree well with the inferred mass
distribution, indicating that the resulting mass profile is not overly sensitive
to the analysis method.

\begin{figure}
\includegraphics[width=3.4in]{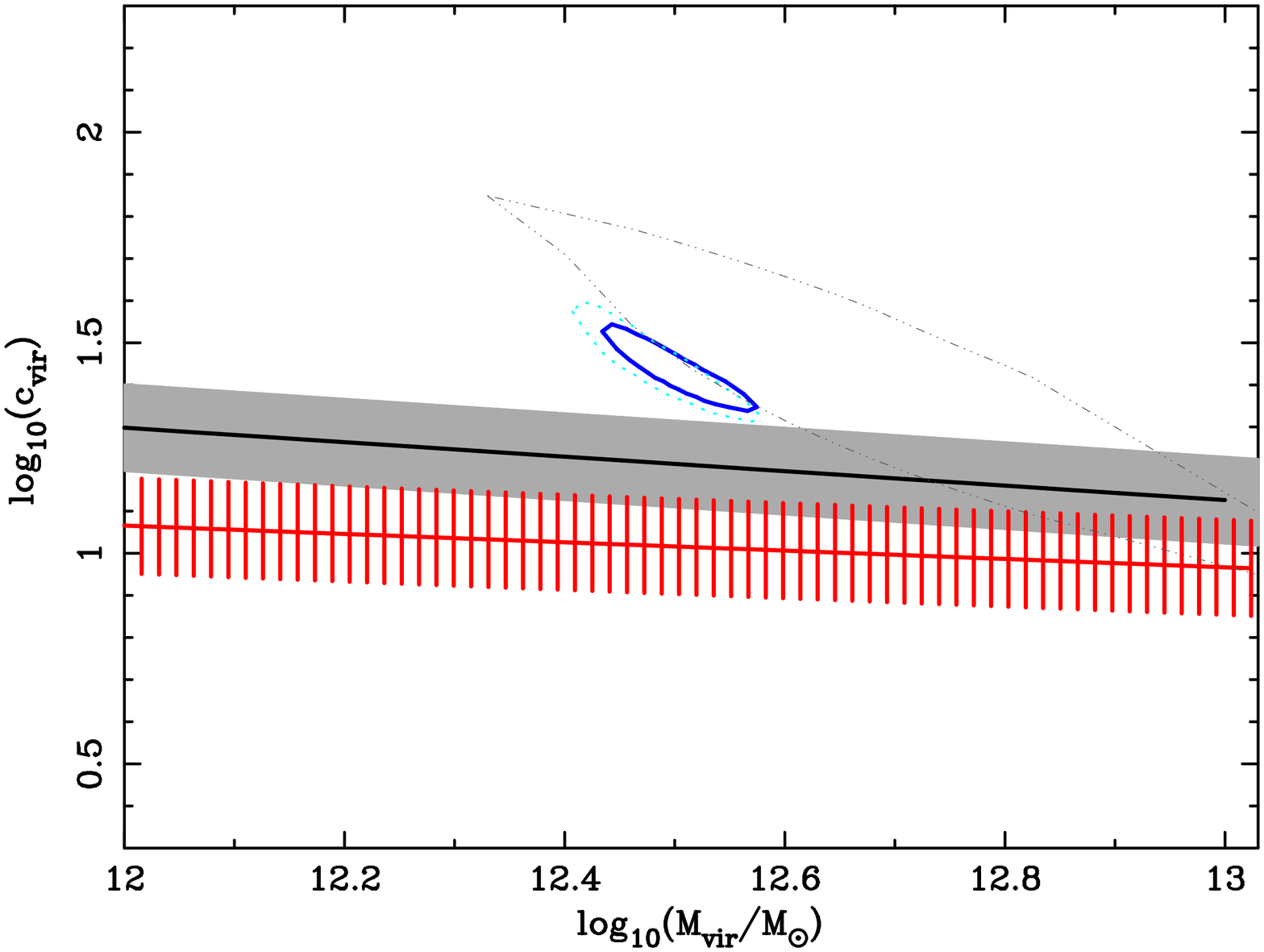}
\caption{Inferred \cvir-\mvir\ relation for \src. We show the marginalized 1-$\sigma$
joint confidence region for \cvir\ and \mvir\ based on our fits (solid blue line). 
The dotted (light blue) curve is the same but for the deprojected data
(\S~\ref{sect_syserr_3d}),
and the dot-dash-dash region is the 1-$\sigma$ error region found by \citetalias{humphrey06a}.
The solid black line and grey shaded region are the mean \cvir-\mvir\ relation, and 
1-$\sigma$ scatter found by \citet{buote07a}, and the red shaded region is the 
theoretical relation for relaxed halos from \citet{maccio08a}.\label{fig_cm}}
\end{figure}
A notable feature of the recovered mass distribution is its flattening outside
$\sim$20~kpc, which allows us to constrain the scale radius, and hence the virial mass
of the dark matter halo. In Fig~\ref{fig_cm}, we show the relation between the 
concentration of the gravitating mass, \cvir,  and the virial mass, \mvir. 
To be consistent with our past work
\citep{buote07a}, the \mvir\ and \rvir\ are derived from the distribution
of the {\em total} gravitating mass, not just the dark matter, and the concentration,
\cvir\ is defined as the ratio of \rvir\ to the characteristic scale of the DM halo.  These results are
consistent with those obtained by \citetalias{humphrey06a} with \chandra, but the constraints are very much
tighter, reflecting both the significantly deeper observation ($\sim$5 times deeper \chandra\ observation,
combined with the data from the very deep \suzaku\ observation) as well as improvements in our mass modelling
procedure. Overlaid on Fig~\ref{fig_cm} we show the theoretical concentration {\em versus} mass relation
from dark matter only simulations \citep[][adopting their ``$\Lambda$CDM1'' relation for 
relaxed halos]{maccio08a}, and the empirical relation from \citet{buote07a}. NGC\thin 720
lies above both relations, but is within the 2-$\sigma$ scatter obtained by 
\citeauthor{buote07a}. The best-fitting marginalized model parameter values and confidence
regions are given in Tables~\ref{table_entropy} and \ref{table_mass}.

\begin{figure}
\includegraphics[width=3.4in]{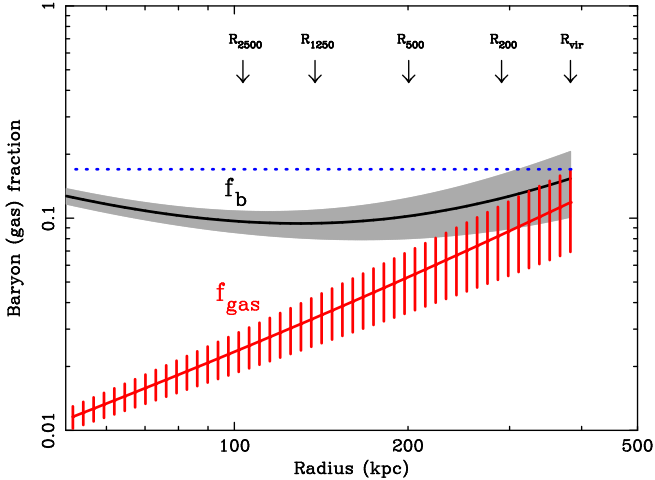}
\caption{Baryon fraction profile inferred from our best-fitting models
(black line). The shaded grey region indicates the 1-$\sigma$ statistical uncertainty
in the fits. The red shaded region indicates the gas fraction profile and 1-$\sigma$
uncertainty. The dotted line indicates the best-fitting Cosmological value of \fb, 
based on the 5-year WMAP data \citep{dunkley09a}, which lies significantly above
the measured \fb\ for most of the radial range. We indicate the physical scales
corresponding to \rvir\ and various other standard radii.\label{fig_fb_profile}}
\end{figure}
\begin{figure}
\includegraphics[width=3.4in]{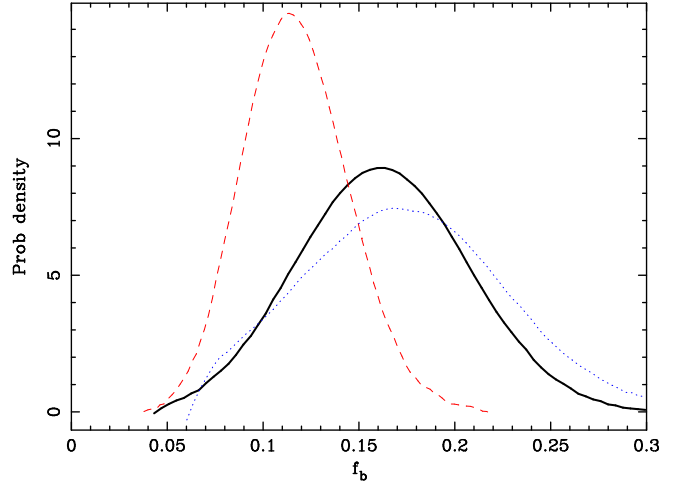}
\caption{Marginalized probability density for \fbvir\ for different choices of the extrapolated 
entropy profile. The solid (black) line is for the default case, where we extrapolate
the measured entropy profile to \rvir. As shown in \S~\ref{sect_entropy}, this extrapolation
is consistent with trends seen in massive galaxy clusters. Also shown are the
cases where we allow the entropy profile to break at $\sim$80~kpc, and the logarithmic
slope ($\gamma$) is much steeper ($\gamma=$2.5; dashed red line) or flatter ($\gamma=0.0$;
blue dotted line). For clarity, we smoothed the distribution functions with a fourth-order 
Savitzky-Golay filter, which spans $\sim$25\%\ of the measured \fb\ range. Although the 
best-fitting values depend to some degree on this choice 
(see also \S~\ref{sect_syserr_entropy}), in all cases \fbvir\ is close to the 
Cosmological value (0.17).
\label{fig_fb_prob}}
\end{figure}
In contrast to our measurement in \citet{humphrey06a}, our adopted modelling procedure
did {\em not} require us to adopt a strong prior on \fb\ in order to obtain 
interesting virial mass constraints. This allows us to measure \fb\ directly 
from our fit. In Fig~\ref{fig_fb_profile}, we show how the enclosed 
baryon fraction and gas fraction vary as a function of radius, and we tabulate
the gas fraction ($f_g$) and \fb, measured at different over-densities,
 in Table~\ref{table_fb}. Outside $\sim$100~kpc 
($\sim$\rtwentyfive), our model predicts that the \fb\ profile rises gently, from 
$\sim$10--15\%.
In Fig~\ref{fig_fb_prob} we show the posterior probability density function for \fbvir\ 
(\fb\ extrapolated to \rvir), {which is tightly constrained to around $\sim$15\%, very
close to the Cosmological value}.
A potentially important source of uncertainty in this extrapolation is the 
shape of the entropy profile. In Fig~\ref{fig_fb_prob} we show the equivalent posterior
probability density functions for two different entropy extrapolations, which should bound all
likely entropy profiles. We discuss these fits in detail in \S~\ref{sect_syserr_entropy}.
Clearly, uncertainties the entropy extrapolation do not
produce a large systematic shift in our inferred \fbvir. 

\section{Systematic error budget} \label{sect_syserr}
In this section, we address the sensitivity of our results to
various data analysis choices that were made, including the
choice of prior. In most cases, it is difficult or impractical to
express these assumptions through a single additional model
parameter over which one might hope to marginalize, and so we
adopted the pragmatic approach of investigating how our results
changed if the assumptions were arbitrarily adjusted. We focused
on those systematic effects likely to have the greatest impact on
our conclusions, and list in Tables~\ref{table_mass} and \ref{table_fb}
the change to the marginalized value of each key parameter for each
test.  We outline below how each test was performed. Those
readers uninterested in the technical details of our analysis may
wish to proceed directly to Section~\ref{sect_discussion}

\subsection{Dark Matter profile} \label{sect_syserr_mass}  \label{sect_syserr_ac}
One of the major sources of uncertainty on the recovered \fb\
is the choice of DM mass model. While our default model (NFW) is 
theoretically motivated, and supported by observations of galaxy
clusters \citep[\eg][]{vikhlinin06b}, we also experimented with a 
so-called ``cored logarithmic'' mass model, {which has been widely used
in studies of elliptical galaxies \citep{binney08a}, including
some recent stellar dynamical studies \citep[\eg][]{thomas07a,gebhardt09a}}.
This model tends to predict higher masses at large radii than NFW, resulting in a significantly larger 
\mvir, and correspondingly a smaller value of \fb\ when extrapolated within \rvir.
Nevertheless, at radii corresponding to higher mass overdensities, 
the extrapolation range is smaller and the discrepancy
is significantly reduced, so that the choice of DM model has little effect
on \fbtwentyfive\ (``$\Delta$DM profile'' in Tables~\ref{table_mass} and \ref{table_fb}). 
{In any case, we note that the cored logarithmic model is less well-motivated
theoretically than the NFW profile. Furthermore, we find that the data slightly
favour the NFW model; although the best-fitting $\chi^2$ is comparable between the 
two cases, the ratio of the Bayesian evidence is $\sim 1.5\times 10^{-4}$, implying
that the cored logarithmic model, with the adopted priors (a flat prior on the 
asymptotic circular velocity, between 10 and 2000~$km\ s^{-1}$, and a flat 
prior on $log_{10} r_c$, where $r_c$ is the core radius, over the range
$0\le log_{10} r_c \le 3$.) is a poorer description of the data at $\sim$3.8-$\sigma$.}

One modification to the (NFW) DM profile shape that is theoretically 
predicted is ``adiabatic contraction'' \citep{blumenthal86a,gnedin04a},
where the DM halo density profile reacts to the gravitational influence of 
baryons that are condensing into stars by becoming cuspier. There is little observational
evidence of this effect (\eg\ \citetalias{humphrey06a}; \citealt{humphrey09d};
\citealt{gnedin07a};  although, see \citealt{napolitano10a}), and it may be 
over-estimated by existing models \citep{abadi09a}, so we did not employ 
it in our default analysis. Nevertheless, we experimented with modifying the 
NFW halo profile to account for this effect with the algorithm described by 
\citet{gnedin04a}\footnote{Using the CONTRA code publicly available from http://www.astro.lsa.umich.edu/$\sim$ognedin/contra/}. The principal effect of this modification was to 
lower slightly the stellar M/L ratio and, consequently, the inferred
\fb\ (``$\Delta$AC'' in Tables~\ref{table_mass} and \ref{table_fb}).

\subsection{Background} \label{sect_syserr_background}
Our treatment of the background is a potentially serious source of 
systematic error. To investigate this, we adopted an alternative 
approach in which we used, 
for the \chandra\ data, the standard blank-field events files 
distributed with the CALDB to extract a background spectrum for each
annulus. Since the blank-field  files for each CCD have different 
exposures, spectra were accumulated for each CCD individually, scaled to a
common exposure time and then added. The spectra were renormalized to match the 
observed count-rate in the 9--12~keV band. These ``template'' spectra were
then used as a background in \xspec, and the background model components were
omitted from our fit. We found that this approach gave a poorer fit to
the data than the full
modelling procedure used by default, and the overall density and temperature
profiles were noticeably affected at large radii.
For the \suzaku\ data, there are no blank-fields files available, but we 
used the standard non X-ray background spectra generated by 
{\em xisnxbgen} tool to remove the instrumental background,
but maintained the sky background model components in our fit. Fitting the
resulting \chandra\ and \suzaku\ temperature and density profiles 
with our mass modelling apparatus, our results were not substantially affected
(``$\Delta$Background'' in Tables~\ref{table_mass} and \ref{table_fb}).

\subsubsection{Solar Wind Charge Exchange} \label{sect_syserr_swcx}
An additional background component can arise from 
the interaction of the Solar wind with interstellar material
and the Earth's exosphere. This should manifest itself as a 
time-variable, soft component that can be modelled as a series
of narrow Gaussian lines, the intensity of which correlate 
with the Solar wind activity \citep[\eg][]{snowden04a}.  
In the interests of minimizing the complexity of the spectral models,
we do not by default explicitly account for this 
so-called ``Solar wind charge exchange'' (SWCX) in our fits.
To some degree this component is degenerate with the sky background,
which we fitted in our modelling, and so this omission
may not be a significant source of bias in our analysis. Nevertheless, it is 
important to confirm that it is not.

To assess the contamination from the SWCX component, we 
first identified periods of low Solar wind proton flux, as measured
by the Solar Wind Electron Proton Alpha Monitor (SWEPAM) 
instrument aboard the Advanced Composition Explorer (ACE) spacecraft 
\citep{mccomas98a}\footnote{Based on the 
publicly released data available
from http://www.srl.caltech.edu/ACE/ASC/index.html}.  
Following \citet{snowden04a}, we assumed the SWCX component
is negligible for a Solar wind proton flux level \ltsim $3\times 10^8 cm^{-2}\ s^{-1}$.
Selecting only times for which SWEPAM measured a proton flux below this limit, 
we were left with $\sim$145~ks of \suzaku\ data. Still, since the proton flux was only
mildly enhanced during the (short) excluded periods, we found that the soft 
(0.5--1.0~keV) X-ray count-rate was only enhanced by a few percent in the entire 
(\ie\ unfiltered) data as compared to the quiescent period.
We found that there was very poor SWEPAM coverage 
of the periods during which the \chandra\ data were taken,
and so we did not attempt to filter the \chandra\ data in the same way. Instead
we modified the sky background component  by adding 
a number of narrow Gaussian lines (fixed at the energies 
reported for the SWCX component by \citeauthor{snowden04a}). 

The corrected \chandra\ and \suzaku\ spectra were 
fitted to obtain the temperature and density profiles, and folded through
our mass modelling apparatus. We found that our results were not substantially
affected by correcting for the SWCX component (``$\Delta$SWCX'' in 
Tables~\ref{table_mass} and \ref{table_fb}).

\subsection{Entropy profile extrapolation} \label{sect_syserr_entropy}
The extrapolation of the entropy profile is one of the 
prime sources of uncertainty in our determination of \fb\ at large scales.
As we will show in \S~\ref{sect_entropy},
our default profile is, in fact, consistent with trends seen in massive
galaxy clusters \citep{pratt10a}, giving us confidence in its use. Nevertheless, 
it is important to explore plausible alternative shapes. 
Motivated by observed entropy profiles in galaxy groups and by 
the monotonically rising entropy profiles required by hydrostatic
equilibrium, we investigated two pathological extrapolations that should
bound any plausible model. {Specifically, we adopted a powerlaw shape
(entropy $\propto R^\gamma$)
outside $\sim$80~kpc, with $\gamma=0$ (\ie\ a flat entropy profile), which
is a secure lower boundary assuming stablity against convection, or 
$\gamma=2.5$. This latter value was chosen arbitrarily, and is far larger
than has been observed in real systems.}
This choice primarily affects the extrapolated temperature, rather 
than the density profile, and so \fb\ does not change substantially
with these choices (``$\Delta$Entropy'' in Tables~\ref{table_mass} and \ref{table_fb}; 
Fig~\ref{fig_fb_prob}).

\subsection{Deprojection} \label{sect_syserr_3d}
In the present work, we opted to fit the projected, rather than the deprojected
data (as done, for example, in \citetalias{humphrey06a}). In general, fitting
the projected data leads to smaller statistical error bars, but potentially larger 
systematic uncertainties \citep[\eg][]{gastaldello07a}, and so it is important to investigate
the likely magnitude of such errors. To do this, we examined the effect on our results
of spherically deprojecting the data. We achieved this by  {using the 
\xspec\ {\em projct} model\footnote{In practice, it was more convenient
to emulate the behaviour of projct by adding multiple ``vapec'' plasma
models in each annulus, with the relative normalizations tied 
appropriately \citep[\eg][]{kriss83a}. This allowed data from the multiple 
\suzaku\ instruments to be fitted simultaneously.}. To account for emission
projected into the line of sight from regions outside the outermost annulus,
we added an apec plasma model to each annulus, with abundance 0.3 (consistent
with the outermost annulus) and the temperature and normalization determined
from projecting onto the line of sight the best-fitting gas temperature and 
density models described in \S~\ref{sect_results_dm}, but considering the models
only outside $\sim$70~kpc.

Although the error-bars were increased when using the deprojected
(rather than the projected) profiles (``$\Delta$Deprojection'' in 
Tables~\ref{table_mass} and \ref{table_fb}), the best-fitting results 
were not significantly changed. In Fig~\ref{fig_mass_profile}, we show
a series of mass ``data-points'' obtained from the deprojected density
and temperature profiles, using the  ``traditional''
mass analysis method described in \citet[][see also 
\citealt{humphrey10a}]{humphrey09d}. These data agree very well with the 
best-fitting mass model found in our projected analysis.
Similarly, the entropy profile derived directly from the 
deprojected data  agrees well with the results of our projected
analysis (Fig~\ref{fig_entropy}).}


\subsection{Priors} \label{sect_syserr_priors}
Since the choice of priors on the various parameters is arbitrary in our analysis,
it is important to determine to what extent they could affect our conclusions.
To do this, we replaced each arbitrary choice in turn with an alternative, reasonable prior.
Specifically, for each parameter describing the entropy profile, we switched from a 
flat prior on that parameter to a flat prior on its logarithm. We replaced the flat
prior on the stellar M/L ratio with a Gaussian prior, the mean and $\sigma$ of which
were determined from the stellar population synthesis model results reported in \citetalias{humphrey06a}. 
We used a flat prior on the DM halo mass, rather than on its logarithm,
 and, instead of the flat prior on $\log c_{DM}$, we adopted the distribution of 
c around M found by either \citet{buote07a} or \citet{maccio08a} as a (Gaussian) prior.
The effect of these choices is no larger than the statistical errors on each parameter,
especially for the baryon fraction measured at $R_{200}$ or higher overdensities 
(``$\Delta$Fit priors'' in Tables~\ref{table_mass} and \ref{table_fb}).

\subsection{Stellar light} \label{sect_syserr_stars}
As discussed in \citetalias{humphrey06a}, careful treatment of the 
stellar light is essential for obtaining an accurate 
measurement of the gravitating mass profile. Since the stellar light
deprojection is not unique, in particular as the inclination angle
is not {\em a priori} known for an elliptical galaxy, we investigated
the associated systematic uncertainty by first varying it
within reasonable limits (\S~\ref{sect_mass}). Specifically,
we lowered it as far as 70$^\circ$, since the highly elliptical projected
isophotes of \src\ make it unlikely to be observed at a much lower
inclination. As a more extreme test of the influence of the stellar
light on our fits, we also experimented with excluding all the data within
the central $\sim$4~kpc, where the stars dominate the mass profile. 
We find that the stellar M/L ratio becomes (unsurprisingly) much more 
uncertain when we adopt this approach, but the best-fitting masses and 
\fb\ were not substantially affected (``$\Delta$Stars'' in 
Tables~\ref{table_mass} and \ref{table_fb}).

\subsection{Emissivity correction} \label{sect_syserr_emissivity}
{In our default analysis, the projected temperature and density
profile are weighted by the gas emissivity,
folded through the instrumental responses
\citep[for details, see Appendix B of][]{gastaldello07a}. Since the
computation of the gas emissivity 
assumes that the three dimensional gas abundance profile
is identical to the projected profile (which is unlikely to be true), 
it is important to assess how
sensitive our conclusions are to the emissivity correction.  To do
this, we adopted the extreme approach of ignoring the spatial 
variation of the gas emissivity altogether. We found that this 
had a very small effect on our results
($\Delta$Emissivity in Tables~\ref{table_mass} and \ref{table_fb}). 
}

\subsection{Remaining tests} \label{sect_syserr_instrument} \label{sect_syserr_covariance} \label{sect_syserr_distance} \label{sect_syserr_radius} \label{sect_syserr_spectra}
We here outline the remaining tests we carried out, as summarized
in Tables~\ref{table_mass} and \ref{table_fb}. First of all, since the 
inter-calibration of the \suzaku\ XIS units may not be perfect,
we experimented with using only one of the units in the \suzaku\ analysis,
and cycled through each choice.
We found that the XIS1 (back-illuminated) unit appeared most inconsistent
with the \chandra\ data, and so also investigated using only units 0, 2 and 3.
We also considered using only the \chandra\ data. While these choices had
a significant effect on \fbvir, we found that \fb\ at higher overdensities
was more resilient (``$\Delta$Instrument'').

To assess the impact of various spectral-fitting data analysis choices on 
our results, in turn we varied the neutral hydrogen column density 
by $\pm$25\%, performed the fit over a restricted energy range (0.7--7.0~keV, or
0.5--4.0~keV), and replaced the APEC plasma model with a MEKAL model. The impact of 
these choices is comparable to the statistical errors on the parameters
(``$\Delta$Spectral'').

To examine the error associated with distance uncertainties, 
we varied the distance to \src\ by the1--$\sigma$ error given in 
\citet{tonry01}. The effect on our results is minor (``$\Delta$Distance'').
To examine the
sensitivity of our fits to the radial range studied, we excluded the data outside
$\sim$40~kpc. This exclusion made the extrapolation more uncertain, especially
for \fbvir\ (``$\Delta$Fit radius'').
Finally, to examine the possible errors associated with our 
treatment of the covariance between the density data-points, we investigated 
adopting a more complete treatment that considers the covariance between all the 
temperature and density data-points, as well as adopting the more standard 
(but incorrect) approach of ignoring the covariance altogether. The effect is not
large (``$\Delta$Covariance'' in Tables~\ref{table_mass} and \ref{table_fb}; Fig~\ref{fig_cm}).

\section{Discussion} \label{sect_discussion}
{The combination of the new, high quality data for \src\ 
and our updated, Bayesian hydrostatic modelling procedure 
has greatly improved our constraints on its gravitating matter and
hot ISM. We here discuss the implications of these new measurements
at some length.}

\subsection{Hydrostatic equilibrium} \label{sect_discuss_he}
The ability of our hydrostatic model to reproduce the temperature and 
density profiles of the gas in \src\ strongly suggests
that the gas is close to hydrostatic; 
despite highly nontrivial temperature and density profiles,
a smooth, physical mass model
and a monotonically rising entropy profile (required for stability
against convection) were able to reproduce them well.
If the hydrostatic approximation is seriously in error, this would 
require a remarkable conspiracy between the temperature and density 
data points. The closeness of the system to hydrostatic 
is unsurprising given its remarkably relaxed X-ray morphology 
\citep[Fig~\ref{fig_images};][]{buote02b,buote96d,buote94} and lack
of substantial radio emission \citep{fabbiano87a}.
Numerical simulations of structure formation suggest that deviations
from hydrostatic equilibrium in morphologically relaxed-looking 
systems are not large, so that errors in the recovered mass 
distribution  should be no larger than $\sim$25\%\
\citep[\eg][]{tsai94a,buote95a,nagai07a,piffaretti08a,fang09a}. 
Such a level of nonthermal support is comparable to the $\sim$10--20\%\
discrepancy found between X-ray and stellar dynamical mass measurements
by \citet{churazov08a} for two galaxies that are manifestly more 
disturbed than \src. 

In the case of \src, there is additional, albeit circumstantial,
 evidence which supports the 
hydrostatic approximation. First, the best-fitting virial mass
is in good agreement with the modest (although uncertain) velocity dispersion of the dwarf
companions about \src, \citep[$117\pm54\ {\rm km\ s^{-1}}$, corresponding
to a virial mass of $4\pm4 \times 10^{12}$\msun:][]{brough06a}.
In our studies of other systems, we have used the 
good correspondence between the measured stellar M/L ratio and that predicted
by single-burst stellar population synthesis (SSP) models that assume a \citet{kroupa01a}
IMF, to infer that there is little nonthermal pressure \citep{humphrey09d}. 
In the case of \src, the measured M/\lk\ ratio ($0.54\pm0.05$) is actually 
{\em higher} than the predictions of the SSP models ($0.35\pm0.07$: \citetalias{humphrey06a}).
A natural way to bring these results into agreement 
is if the young \citep[$\sim$3~Gyr:][]{humphrey05a,rembold05a} stellar population
that dominates the light only represents recent star-formation superimposed
on an older, underlying stellar population with a higher M/\lk\ ratio. Such 
a picture is supported by the multiple-aged stellar population inferred by 
\citet{rembold05a}. We note that the measured M/\lk\ value could also be 
reconciled with the predictions of SSP models if a Salpeter IMF 
is adopted  (for which M/\lk$\sim$0.54; taken in the context of our 
previous work this would indicate a non-universal IMF). 
Alternatively, if AC operates, making \src\ unique amongst the systems we have studied in
the X-ray, the measured M/\lk\ would be $0.39\pm0.04$, also reconciling the 
fit result and the SSP predictions.
While it is evidently difficult to draw
strong conclusions based on the M/\lk\ ratio, none of these arguments
indicates that the measured M/\lk\ is obviously underestimated. This is important
since most models predict that non-hydrostatic effects lead to an {\em underestimate} of the mass
\citep[\eg][]{nagai07a,churazov08a,fang09a}.

\citet{humphrey10a} studied \src, as part
of a sample of galaxies, groups and clusters, finding that the total
gravitating mass profile in the central region (\ltsim 30~kpc) is 
close in shape to a singular isothermal sphere distribution. This shape is 
what is approximately inferred from lensing plus stellar dynamics 
studies of similar galaxies 
\citep[\eg][]{koopmans09a}. Therefore, any deviations from hydrostatic
equilibrium are unlikely to have substantially affected the overall 
shape of the gravitating mass profile, at least over this region. 
Nevertheless, it is plausible that a constant fraction of the pressure
over this region comes from nonthermal effects, which would just reduce
the overall normalization of the measured mass model. Such a scenario
would imply that the true \rvir\ is underestimated, while the scale
radius is approximately correct; hence \cvir\ would be underestimated,
putting it in more tension with theory. Conversely, it is unlikely that deviations
from hydrostatic equilibrium could lead to the scale radius also 
being underestimated, since that would require the nonthermal pressure
support to increase dramatically outside $\sim$20~kpc. Since \src\ is 
an isolated system, it is difficult to imagine what processes 
could give rise to such an effect.

One effect which could produce deviations from hydrostatic equilibrium
at the smallest scales is gas rotation induced by spin-up of subsonically 
inflowing gas.
Considering the very relaxed galaxy NGC\thin 4649 \citet{brighenti09a}
demonstrated that angular momentum
conservation (if there is appreciable rotation in the stars) flattens 
the X-ray ellipticity significantly in the centre of the system (where deviations from
hydrostatic equilibrium are also most significant). \src\ exhibits sufficient
major-axis rotation \citep{binney90a}, but only a hint of a central ellipticity
rise within the innermost $\sim$1~kpc \citep{buote02b}. Since our results
are not very sensitive to the exclusion of data within the central $\sim$4~kpc
(\S~\ref{sect_syserr_radius}), it seems unlikely that this effect could have 
an impact on our results, even if it is operating in \src.

{Our conclusions on the state of hydrostatic equilibrium in  \src\ differ significantly 
from those of \citet{diehl07a}. Those authors investigated the X-ray and optical isophotal  
ellipticities at a fixed, small scale in an heterogenous sample of early-type galaxies,
concluding that the lack of a correlation between them (in conflict with would be expected if  hydrostatic
equilibrium holds exactly) indicates that hydrostatic equilibrium is not ubiquitous and, therefore,
should never be assumed (even approximately) in mass analysis (and, in particular, for \src). 
We consider their conclusions, however, to be extreme and misleading.
Firstly, their method does not allow them to say anything at all about individual
objects, and only indicates that  hydrostatic equilibrium is unlikely to hold {\em perfectly}
in a galaxy randomly drawn from their sample, provided it is 
{\em chosen without any consideration of its morphology}.
Since their sample contained galaxies with such large-scale asymmetries that we would
not recommend the routine application of hydrostatic mass methods
(\eg\ NGC\thin 4636: \citealt{jones02a} and M\thin 84: \citealt{finoguenov01}),
and since they focused on the smallest scales 
(where X-ray point-source removal is most challenging 
and where AGN-driven disturbances are most serious), in contrast to the large
scales that are most important for dark matter analysis, the implications
for morphologically relaxed systems such as \src\ or NGC\thin 4649 \citep{humphrey08a}
are unclear.
Secondly, even if it does not hold {\em perfectly}, hydrostatic equilibrium
can still be a useful approximation. With their hydrodynamical models, \citet{brighenti09a}
reconciled the X-ray and optical ellipticity profiles of NGC\thin 4649 by requiring modest
gas motions, but the gas remained very close to hydrostatic (especially outside the innermost
$\sim$1~kpc).
 Substantial non-hydrostatic  gas motions are likely to manifest themselves 
as  sharp features in the surface brightness and temperature profiles of the hot gas,
for example those seen in the core of M\thin 87 \citep{forman07a,million10a}. 
Still, even for M\thin 87,
\citet{churazov08a} found that the gas was close to hydrostatic away from the regions of 
most significant disturbance. In the case of \src, the profiles are significantly smoother
than in M\thin 87 (Fig~\ref{fig_profiles}), while the hydrostatic isophotal analysis undertaken
by \citet{buote02b} found evidence for a dark matter halo ellipticity in excellent agreement
with the predictions of cosmological models (implying the gas is not grossly out of 
equilibrium). In summary, in our detailed analysis of \src\ 
and other systems \citep{humphrey09d,brighenti09a}, we find no evidence supporting 
\citeauthor{diehl07a}'s  contention that hydrostatic equilibrium is a very poor
approximation in morphologically relaxed galaxies
(which seems little more than ``guilt by association'').}

\subsection{Baryon fraction}
\subsubsection{Robustness of \fb\ measurement}
Based on our self-consistent hydrostatic mass model for \src, and the fit to the density
profile, we were able to obtain unprecedented \fb\ constraints in a galaxy-scale system
at large radius.  Since the current X-ray data only reach $\sim$0.7\rtwentyfive\ 
(with the field of view of \suzaku\ being the primary limiting factor), most of 
the constraints are
based on an extrapolation. Nevertheless, this is done in a well-motivated, self-consistent manner,
relying only on how the mass and entropy profiles behave at large radius, and the validity
of the hydrostatic approximation at large scales. Of these, the dark matter halo 
mass profile has by far the biggest
influence on the inferred \fb, as discussed in \S~\ref{sect_syserr_mass}. Specifically,
\fbvir\ changes from $\sim$0.15 when an NFW dark matter halo is adopted to $\sim$0.04
for a ``cored logarithmic'' dark matter mass model, which  mostly reflects the higher 
\mvir\ in the latter case. While this is a concern, the NFW model is much better
motivated cosmologically,  and supported observationally by high-quality 
X-ray and lensing studies of other (albeit more massive) systems 
\citep{lewis03a,pointecouteau05a,vikhlinin06b,gavazzi07a}. Still, even if the true DM 
mass model
differs unexpectedly from NFW, we find that \fb\ measured within regions of larger
overdensity (\eg\ $R_{2500}$) is much less sensitive to this issue, reflecting the 
smaller range of extrapolation.

{While the extrapolation of the entropy profile in our analysis is also a potential cause
for concern, this choice does not significantly  affect our conclusions (Fig~\ref{fig_fb_prob}). 
In fact, even allowing for a pathological variation in the slope of the entropy 
profile at large radius ($\gamma$ ranging from 0--2.5), we find that the best-fitting
\fbvir\ varies by only $\sim$1-$\sigma$. 
This apparent lack of sensitivity to $\gamma$ arises because changing the entropy 
distribution in our hydrostatic models tends to affect the implied temperature more strongly than
the gas density.
}

Another factor which may complicate the extrapolated \fb\ values is the hydrostatic 
approximation. As we argued in \S~\ref{sect_discuss_he}, there is good evidence to
suggest the gas is hydrostatic within $\sim$75~kpc, but we cannot directly assess the validity 
of this approximation
at larger scales. Most mechanisms that can produce strong deviations from the single-phase, 
hydrostatic approximation in an isolated galaxy would actually 
provide non-thermal support to the gas \citep[\eg][]{zappacosta06a,nagai07a,churazov08a,fang09a}.
This would produce a {\em flatter} gas density profile than the purely hydrostatic 
case, thereby {\em increasing} the actual baryon fraction over that inferred from
our extrapolation. (An exception is if the gas is globally outflowing, in which case
the gas is over-pressured, and the hydrostatic model would over-estimate \fb. 
However, such a global outflow would have to be triggered
by feedback from the central galaxy, so it seems implausible for there to be an outflow
outside $\sim$75~kpc, but no evidence of one at smaller scales.) Numerical cosmological simulations
of hot halos around galaxies do, indeed, predict that they are
quasi-hydrostatic \citep[\eg][]{crain10a}.

Our measurement of \fb\ given in Table~\ref{table_fb} does not 
include all of the baryons within the system, since it ignores the 
dwarf companions. In practice, the total K-band luminosity of the detected dwarf galaxies 
is only $\sim$10\%\ of that of \src\ \citep{brough06a}, and the HI content 
of these galaxies is much smaller still \citep{sengupta07a,kilborn09a}. If the luminosity 
function of these galaxies is similar to that of Milky Way dwarfs, their combined
luminosity will be dominated by the brightest few members \citep{tollerud08a}, and so the baryons
hosted by as yet undetected, fainter dwarfs will not significantly affect this estimate.
Additional baryons could plausibly exist in a hard-to-detect extended stellar envelope, 
analogous to the ``intra-cluster light'' seen in clusters and massive groups, but galaxy
formation models suggest that, for a system of this mass, at most $\sim$10\%\ of the total
stellar mass could be distributed in this way \citep{purcell07a}. Finally, we note that 
\src\ itself does not contain significant cool gas  
\citep{huchtmeier94a,welch10a}.
Failing to include these uncounted
sources of baryons, therefore, may bias our inferred \fb\ values low by at most $\sim$0.01.

\subsubsection{Implications}
\begin{figure}
\centering
\includegraphics[height=4.5in]{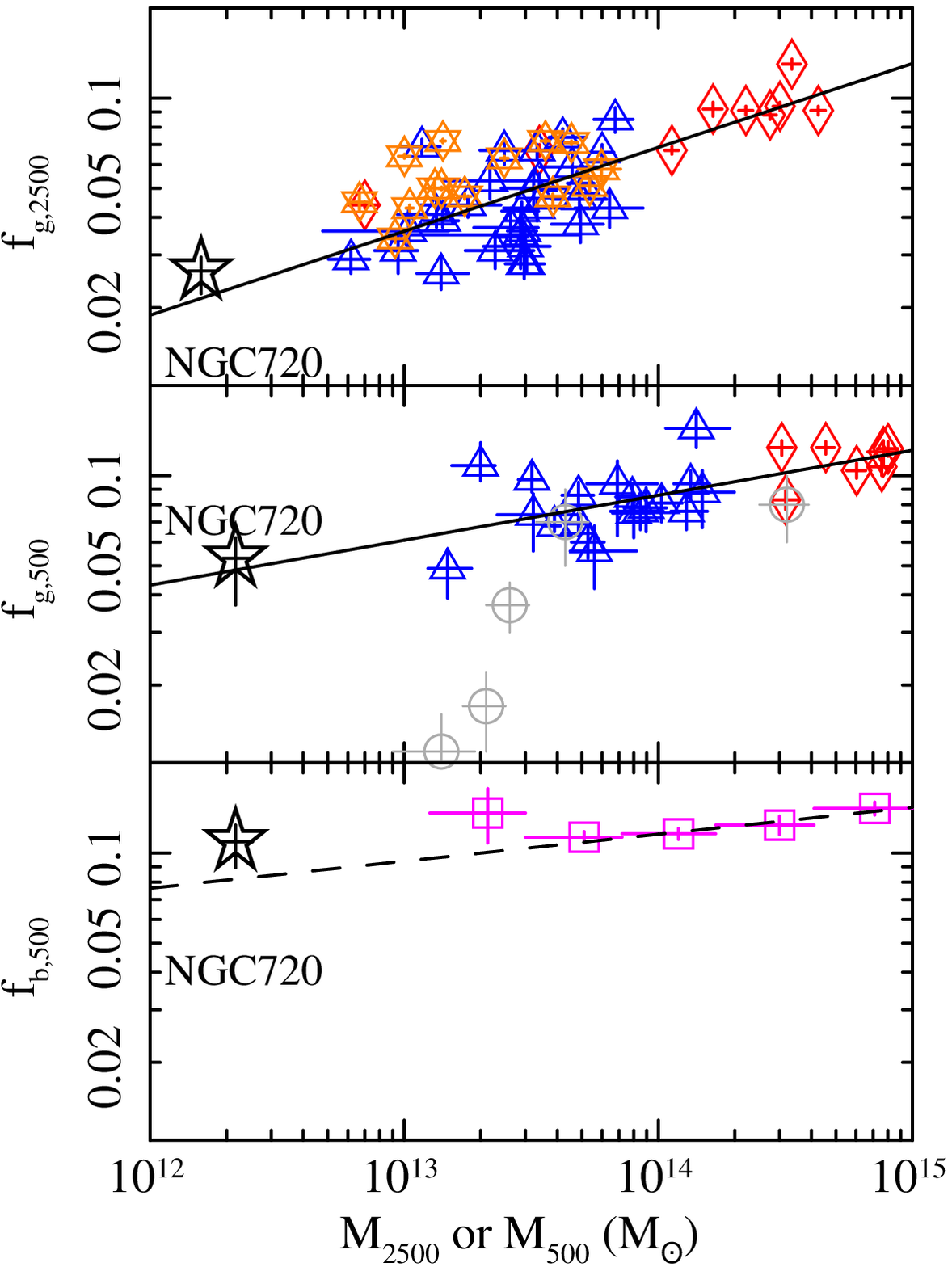}
\caption{{{\em Top panel:} Gas fraction within \rtwentyfive\ shown as a function
of \mtwentyfive\ for the isolated galaxy \src\ (five-pointed star) and a literature sample of groups
and clusters (\citealt{vikhlinin06b}: diamonds; \citealt{gastaldello07a}: six-pointed
stars; \citealt{sun08a}: triangles). The best powerlaw fit to the data is shown as a solid
line. The gas fraction within \rtwentyfive\ of \src\ is clearly consistent with an 
extrapolation of the trend for galaxy groups and clusters.
{\em Centre panel:} The same, but within an overdensity of 500. We overlay additional
literature data from \citet[][circles]{dai09a}.}
{\em Lower panel:} Baryon fraction within \rfive\ as a function of \mfive\ for 
\src\ and the mean data for the group and cluster sample of \citet[][squares]{giodini09a}. The 
best-fitting powerlaw relation from \citeauthor{giodini09a} is overlaid (dashed line).
\src\ is clearly consistent with an extrapolation of this relationship to low masses.
\label{fig_fgas_v_m500}}
\end{figure}
Based on our hydrostatic mass models,
the total {\em stellar mass fraction} within \rvir\ is small
($\sim0.035\pm0.003$), but in good agreement with that generally inferred from gravitational
lensing studies of early-type galaxies \citep{hoekstra05a,gavazzi07a}. Most of the baryons
within \rvir\ are, however, in the form of the hot gas. Including both gas and stars,
the measured \fb\ profile
is relatively flat at $\sim$0.10--0.15 between \rtwentyfive\ and \rvir\ 
(Fig~\ref{fig_fb_profile}), becoming completely consistent with the Cosmological
value \citep[0.17:][]{dunkley09a} when extrapolated to \rvir. 
In \citetalias{humphrey06a} we reported a lower value of \fb\ within \rvir\ for \src\ 
($\sim$0.04), based on
simpler hydrostatic model fits to data from a much shallower \chandra\ observation. That constraint, however,
was almost entirely driven by a strong prior that imposed a tight correlation between \mvir\
and \fb; when relaxing this prior, we found that \fb\ was very poorly constrained by those data.

{A number of authors have investigated the gas (or baryon) fraction in galaxy 
groups and clusters, typically finding a shallow dependence on the virial mass of the 
halo. In the upper panel of Fig~\ref{fig_fgas_v_m500} we show the gas fraction
measurements within \rtwentyfive\ ($f_{g,2500}$) {\em versus} \mtwentyfive\ for the group and cluster samples 
of \citet{vikhlinin06b}, \citet{gastaldello07a} and  \citet{sun08a}\footnote{For clarity, we omit systems
with large error-bars ($\sigma(f_g)$\gtsim 0.01). Where the samples overlap, we use the results,
in order of preference, from \citet{gastaldello07a} and \citet{sun08a}.}. Overlaid is the locus of \src,
which is in good agreement of an extrapolation of the powerlaw trend inferred for 
the more massive systems. Fitting a model of the form
\begin{equation}
\log_{10} f_{g} = A + B \log_{10} \left( \frac{M}{2\times 10^{14}M_\odot}\right)  \label{eqn_linear}
\end{equation}
to these data, using a method that takes into account errors on both axes and intrinsic
scatter in the y-direction \citep[specifically, that employed in][]{humphrey08b},
we found $A=-1.08\pm 0.03$ and $B=0.28\pm 0.04$, with an intrinsic scatter of 0.11~dex.
In the centre panel of Fig~\ref{fig_fgas_v_m500},
we show a similar comparison between $f_{g,500}$ and \mfive, similarly finding that the 
locus of \src\  is consistent
with an extrapolation of the trends for the massive systems. Fitting a log-linear 
regression
line (Eqn~\ref{eqn_linear}), we obtained $A=-1.02\pm0.02$ and $B=0.15\pm0.03$, 
with an intrinsic scatter of 0.06~dex, in good agreement 
with a fit by \citet{giodini09a} to a 
similar group and cluster sample, but omitting the \src\ datapoint. 
Since the temperature varies only by $\sim$1 order of magnitude for the almost
3 orders of magnitude mass range between the galaxies and clusters,
hydrostatic equilibrium generally requires that massive systems have 
 a more rapidly declining gas density profile than less-massive objects,
and hence a more centrally peaked gas mass distribution,
as compared to the distribution of dark matter.
Given this trend, it is not surprising that
lower-mass objects have a smaller fraction of their total gas mass within \rtwentyfive,
as indicated by the dependence of B on the overdensity.}

Also shown in Fig~\ref{fig_fgas_v_m500} (centre panel) are a set of data-points taken from
\citet{dai09a}, who fitted stacked Rosat All-Sky Survey (RASS) data of 
optically selected galaxy groups and clusters. These points appear quite discrepant
with this relation (and the other data) below $\sim 3\times 10^{13}$\msun,
where they fall sharply. This may indicate real differences in the properties 
of optically selected groups from those selected from X-ray surveys, 
but it may also reflect the significant systematic errors that are involved with
stacking the RASS data \citep[see][]{dai07a}.

While the gas is the dominant baryonic component for the most
massive clusters, the stellar-to-gas mass ratio is actually a strong function of 
\mfive. In \src, approximately half of the baryons within \rfive\ 
are in the stars. Accounting for both gas and 
stars \citet{giodini09a} obtained average \fb\ constraints within \rfive\
for a galaxy group and cluster sample. In the lower panel of Fig~\ref{fig_fgas_v_m500},
we show their data as a function of \mfive, along with the best powerlaw
fit to their data, and the locus of \src. It is immediately clear that \src\
is in good agreement with the extrapolation of the (shallow) trend they observe 
to low masses. \citet{dai09a} also reported total baryon fractions 
for their galaxy group and cluster sample, finding rather lower values of 
\fbtwo\ in their lowest mass systems (\fbtwo $\sim$0.04 at \mtwo 
$\sim 3\times 10^{13}$\msun). This is significantly below \fbtwo\ observed in
\src\ (which has \mtwo\ an order of magnitude lower), but is comparable to its stellar 
mass fraction. Still, as discussed above, the gas fraction estimates from
their work appear discrepant with studies of X-ray selected groups and this
may explain the  inconsistency.

Our study of \src\ clearly demonstrates that it is possible for a system with
\mvir\ as low as a few times $10^{12}$\msun\ to retain a massive, quasi-hydrostatic
hot halo. The total baryon fraction, when extrapolated to the virial radius,
is consistent with the Cosmological value, indicating that massive dark
matter halos around galaxies could be a significant reservoir for some
fraction of the  ``missing baryons'' \citep[\eg][]{fukugita06a}.
The biggest spiral galaxies are believed to reside in halos of comparable mass,
making it particularly interesting to compare the hot halo around \src\ 
to those predicted around spiral galaxies. 
Intriguingly, the total gas content of the system (gas fraction within \rtwo\ of 
$0.08\pm0.03$) does agree well with recent numerical disk galaxy simulations at this
mass range \citep[\eg][]{crain10a}. In these
simulations, most of the gas is in the hot phase, suggesting that the cool baryon content
of the most massive spiral galaxies (\ie\ considering only stars and cool gas) should
be systematically lower than \fb\ measured in \src.

\citet[][see also \citealt{dai09a}]{mcgaugh10a} has assembled \fbfive\ estimates
for a sample of massive disk galaxies, {\em omitting any putative hot halo from the estimate}.
Indeed, \fbfive\ in these systems does  lie slightly below that in \src.
Still, the stellar mass fraction in these spiral galaxies is substantially
higher than in \src, and the modest discrepancy in \fbfive\ leaves
little room for a massive hot halo around the spirals. This appears 
inconsistent with the simulations and,
at face value, might suggest very different star formation efficiencies 
between the different types of galaxy. One difficulty with this 
interpretation, however, is the young ($\sim$3~Gyr) dominant stellar 
population in \src, suggesting a recent formation epoch, possibly
from a merger event involving massive disk galaxies \citep{rembold05a}.
Since the gas fraction can only be reduced during a merger (as it may be
ejected from the galaxy, funnelled onto the central black hole or 
converted into stars), this implies there must have been at least 
as much gas as stars within \rfive\ in the progenitors.
Even allowing for stellar mass loss over the last $\sim$3~Gyr, this would
require similarly inefficient star formation in the precursor
(spiral?) galaxies. We caution that the \mfive\
values for the spiral galaxies in \citeauthor{mcgaugh10a} were obtained 
by scaling the rotation velocity, rather than formal modelling, which 
could plausibly introduce some uncertainties into both \mfive\ and 
\fbfive\ reported in that work.

Finally, it is interesting to compare our results to measurements of \fb\ in the 
Milky Way, around which there is indirect evidence for an extended, hot halo.
This system has \mvir\ in the range $\sim 1$--$3\times 10^{12}$\msun\
\citep{klypin02a,sakamoto03a}, close to that of \src. Based on the
baryonic mass estimate of \citet{flynn06a}, which does not consider any
putative hot halo, \fbvir\ lies in the range $\sim$0.02--0.06.
This is suggestively below our measurement for \src\ and the 
discrepancy is most acute towards the more massive end of the Milky Way
mass range (\ie\ almost the same mass as \src). 
\subsection{Dark Matter Halo}
{We have obtained a high-quality measurement of 
the gravitating mass profile for the isolated elliptical galaxy \src\ between
$\sim$0.002\rvir\ and 0.2\rvir\ ($\sim$0.7\rtwentyfive).
As found in past studies of the hot gas around galaxies
\citep[\eg][]{buote02b,humphrey06a,humphrey09d,fukazawa06a,osullivan07a,zhang07a,nagino09a}, the 
data are inconsistent with a constant M/L ratio model. 
Given the high-quality constraints on the mass profile in \src, 
this means that dark matter is required at 19.6-$\sigma$.}

With these data from deep \chandra\ and \suzaku\ observations, we have been able to obtain
unprecedented constraints on the 
virial mass and dark matter halo concentration at this 
mass range (\mvir $=3.1\pm0.4 \times 10^{12}$\msun). 
Such a low mass (only a few times more massive than the Milky Way)
is consistent with our picture of this system as an isolated 
elliptical galaxy, rather than a galaxy group {\em per se}.
These results are in agreement with, but
very much tighter than, those found by \citet{humphrey06a}
for the same system. This improvement 
partially reflects the much better data
used here ($\sim$100~ks of \chandra\ data, plus $\sim$180~ks of
\suzaku\ time, as compared to the $\sim$17~ks \chandra\ dataset 
used by \citeauthor{humphrey06a}), but is also on account of 
various refinements made to our mass modelling procedure in the 
interim. Of particular significance is the adoption of 
parametrized models for the entropy (rather than the temperature)
distribution, allowing us to enforce a monotonically rising entropy
profile (\ie\ the Schwarzschild criterion for stability against convection).
This efficiently eliminates
unphysical portions of parameter space. Similarly, explicitly
requiring the gas to remain approximately hydrostatic out to the virial radius also
restricts parameter space to regions containing only physical models, whereas 
the fitting procedure of \citetalias{humphrey06a} was less restrictive.

In large part, the tight constraints on the virial mass and concentration are due
to the ability of the data to constrain the flattening
of the mass profile outside $\sim$20~kpc. This scale is much larger
than the effective radius of the galaxy (3.1~kpc), and is 
thus unrelated to the stellar-dark matter
conspiracy \citep{humphrey10a}, and instead must indicate the scale radius
of the dark matter halo. As discussed in \citet{gastaldello07a},
the ability to constrain the virial quantities is very sensitive to the 
accurate measurement of the scale radius. This flattening is not
model dependent; when fitting the  ``cored logarithmic'' DM 
model, we similarly found a flattening in the total gravitating mass
profile at this scale\footnote{We note that the best-fitting mass profile 
agrees with that reported by \citet{humphrey10a}, who found that, within
$\sim$30~kpc, it is approximately consistent with a powerlaw. In that case,
the data lacked the radial coverage to constrain the subtle flattening
between $\sim$20--30~kpc, within which there is only a single data bin.
Furthermore, the slope of the mass profile
was insensitive to the exact radial fitting range adopted, so our conclusions
in \citet{humphrey10a} are not affected by our failure to correct for this 
flattening.}. 

The small DM scale radius in \src\ indicates a high 
concentration;  indeed, we find that it lies 
$\sim 3$ times the intrinsic scatter above the predicted
median relation found in dark matter only simulations for the ``$\Lambda$CDM1''
(``concordance'') cosmology by \citet{maccio08a}.
\citet{buote07a} measured the c-\mvir\ relation empirically 
for a sample of early-type galaxies, groups and clusters, finding
higher concentrations in the galaxy regime 
(\ltsim $10^{13}$\msun) than \citeauthor{maccio08a} Still,
we  find that NGC\thin 720 lies $\sim$2 times the intrinsic scatter 
above this relation\footnote{Strictly speaking, this comparison is 
not independent, since NGC\thin 720 was one of the systems used to
determine the \citet{buote07a} relation, albeit with much poorer
data. Nevertheless, NGC\thin 720 does not appear to be unusual in
comparison to the other systems used by these authors, 
and so it is unlikely that this relation is largely driven by this
one data-point.}. 
The extreme isolation of NGC\thin 720
suggests an early epoch of formation (such that the brightest
sub-halos have merged to form the central object), which should
give rise to a higher DM halo concentration, and may provide a 
possible explanation for the measured \cvir\
\citep[\eg][]{zentner05a}. The very young 
($\sim$3~Gyr) mean age for the central stellar population may 
simply reflect stars formed in a recent gas-rich merger rather than
a realistic estimate of the age of the population as a whole 
(see discussion in \S~\ref{sect_discuss_he}).

One factor which could lead to an over-estimate of \cvir\
is a failure to model correctly the stellar mass component 
\citep{mamon05a,humphrey06a}. If the young stellar population
which dominates the light is, indeed, only a fraction of the 
total population in the galaxy, it is plausible that the assumption
that mass follows light is not correct \citep[although the modest
colour gradient of the galaxy is not consistent with a dramatic
violation of this assumption:][]{peletier90a}. However, excluding
the central $\sim$4~kpc ($\sim$1.3\reff) does not significantly
affect the concentration (\S~\ref{sect_syserr_radius}).
Alternatively, if adiabatic contraction operates, it should increase
the cuspiness of the central DM halo profile, for a fixed 
stellar M/L ratio, which could also lead to \cvir\ being 
over-estimated if it is not accounted for \citep[\eg][]{napolitano10a}.
Fitting an AC model (\S~\ref{sect_syserr_ac}) also led to a 
slightly less concentrated halo, but the effect was not large enough
to reconcile the best-fitting value with the \citet{maccio08a}
predictions.
This comparative  lack of sensitivity of the concentration to the 
details of the stellar modelling is not altogether surprising, since \reff\
(3.1~kpc) is far smaller than the scale radius of the DM halo.



\subsection{Entropy profile} \label{sect_entropy}
\begin{figure}
\includegraphics[width=3.4in]{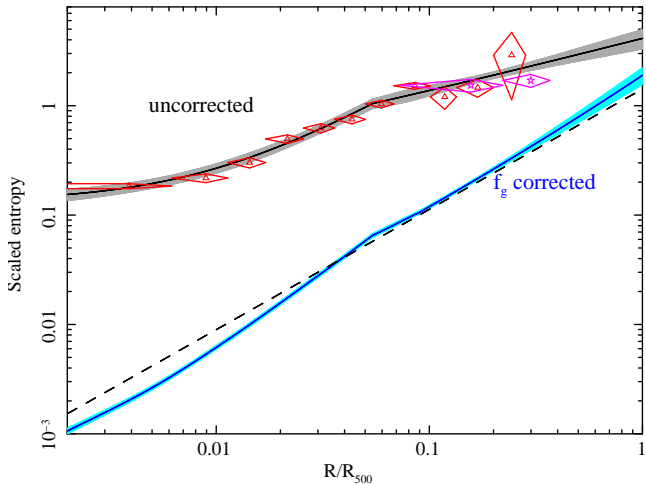}
\caption{1-$\sigma$ confidence region for the entropy profile model of \src, 
scaled by its characteristic entropy
\kfive\ (=27.9 keV cm$^2$), and shown as a function of  \rfive. 
We overlay the deprojected entropy data-points 
(shown in Fig~\ref{fig_entropy}),
similarly scaled. The dotted line indicates the ``baseline'' prediction from gravitational
structure formation \citep{voit05b}. Also shown is the scaled entropy model, corrected
for the gas fraction profile (``$f_g$ corrected''; see text), which agrees very well
with the baseline model. \label{fig_scaled_entropy}}
\end{figure}
\begin{figure}
\includegraphics[width=3.4in]{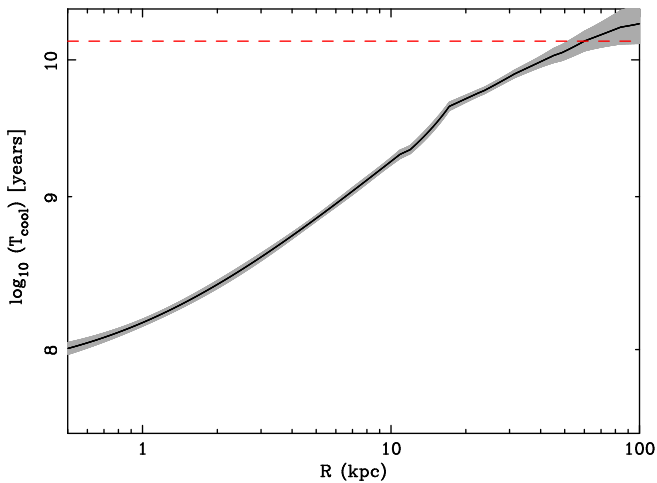}
\caption{Cooling time of the gas (solid line), and its 1-$\sigma$ confidence region
(shaded region). Error bars do not incorporate abundance errors (which are 
less than $\sim$20\%). Gas inside $\sim$60~kpc has a cooling time shorter than the 
Hubble Time (shown as a dashed line).\label{fig_tcool}}
\end{figure}

As expected for gas which is approximately hydrostatic, we obtain a good fit to the 
temperature and density profiles with a monotonically rising entropy profile.
Although the entropy profile slope  in the inner part of the system ($\beta_1$) 
is close to the canonical 1.1 (Table~\ref{table_entropy}) predicted by
gravitational structure formation simulations \citep[considering only gravity
without cooling or feedback:][]{voit05b}, the profile 
flattens at smaller and larger radii. This is consistent with two of the three galaxies
studied in \citet{humphrey09d}, and similarly shaped profiles have also been
observed in galaxy groups \citep[\eg][]{jetha07a,finoguenov07a,gastaldello07b,sun08a,flohic10a}. 

In Fig~\ref{fig_scaled_entropy} we show the 
entropy profile as a function of \rfive, and scaled by the ``characteristic entropy'',
\kfive\ \citep[\eg][]{pratt10a}. Clearly the profile is significantly flatter than
the ``baseline'' model predicted by gravity-only simulations \citep{voit05b},
and the normalization is significantly enhanced, reflecting the injection of entropy by
non-gravitational processes, especially in the inner regions. In comparison with
galaxy groups and clusters \citep[\eg][]{pratt10a,sun08a}, the entropy profile is more
discrepant with the baseline model, reflecting the greater impact of nongravitational
heating on a lower mass halo. Nevertheless, the entropy injection still
appears insufficient to evacuate a significant fraction of the baryons from the halo,
as indicated by the approximate baryonic closure within \rvir. While this may reflect a 
finely-balanced heating mechanism in the system, we note that most of the gas lies at large
radii, where the entropy profiles are much less discrepant 
(for example, $\sim$65\%\ of the gas lies outside \rfive).

If nongravitational processes primarily redistribute
the gas, rather than raising its temperature, we might expect the entropy profile to
be simply related to the baseline model and the gas mass profile. Indeed \citet{pratt10a}
demonstrated with their cluster sample 
that the product $S(R) \left( f_g(R)/ f_{b,U}\right)^{2/3}$, where S is the 
entropy profile, $f_g(R)$ is the gas mass fraction profile (Fig~\ref{fig_fb_profile}) 
and $f_{b,U}$ is the Universal baryon fraction (0.17), is very close to the baseline model.
It is interesting to investigate whether this also holds for lower-mass halos, where the 
impact of nongravitational heating is systematically larger. We show
the ``$f_g$ corrected'' entropy profile for \src\ in Fig~\ref{fig_scaled_entropy}, 
finding that it also lies very close the gravity-only predictions, even in a $\sim$Milky
Way-mass halo. {We note that the corrected entropy profile agrees with this relation 
even when extrapolated outside the projected radii at which there are data, which provides support
for our extrapolated entropy profile.}

{In order to interpret the impact of non-gravitational processes on the 
entropy distribution, it is helpful to consider the cooling time of the gas. In
Fig~\ref{fig_tcool}, we show that, within $\sim$60~kpc, it is shorter than the 
Hubble time. Nevertheless, only $\sim$6\%\ of the gas is currently within this region,
implying that most of the gas in \src\ will not have had sufficient time to cool.
In part this reflects the significant non-gravitational heating that has raised
the entropy even in these outer regions significantly above the baseline model
(Fig~\ref{fig_scaled_entropy}). The average stellar age in \src\ is young 
($\sim$3~Gyr), and if this reflects the time since 
\src\ formed, only the gas within $\sim$15~kpc would have had time to cool,
within which there is presently only $\sim$0.5\%\ ($\sim 10^9$\msun)
of the total gas mass.}

\subsection{Abundance gradient} \label{sect_abundance_gradient}
One of the intriguing results from our study is the detection, with both 
\chandra\ and \suzaku, of a strong negative abundance gradient in the hot gas of \src\
(Fig~\ref{fig_profiles}), making it arguably the lowest-mass system in which
such a gradient has been definitively detected. Although the metallicity of the ISM in early-type
galaxies with intermediate to high X-ray luminosities is now known to be $\sim$Solar,
and comparable to the stellar metallicity \citep{humphrey05a,ji09a},
simple chemical enrichment models typically predict substantially more metals in the 
central part of the halo, where enrichment from supernovae and stellar mass-loss
is concentrated \citep[\eg][]{pipino05a}.
Solutions to this problem may involve large-scale
mixing due to AGN-ISM interaction, or the assembly of structure through mergers
\citep{mathews04a,cora06a,cox06a}. Clearly the spatial distribution of the metals 
in the hot gas is a useful constraint on any model predicting large-scale mixing.

While, to date, very little is known about the radial dependence of \zfe\ in
the ISM of galaxy-scale halos, recent \chandra\ and \xmm\ 
studies of relaxed galaxy groups have revealed similar, centrally peaked
\zfe\ profiles \citep[\eg][]{humphrey05a,buote02a,buote03b,rasmussen07a}.
\citet{mathews04a} were able to explain the overall shape of these group-scale 
profiles by invoking large-scale flows driven by low-level AGN heating.
It remains to be seen whether such a model could also explain the observed 
abundance profile of \src\ which, residing in a much shallower potential well,
should be more susceptible to the effects of feedback.

\section{Conclusions}
Based on an analysis of data from deep new \chandra\ and \suzaku\ observations, we have obtained 
unprecedented constraints on the temperature, density and heavy metal abundance in the 
hot gas halo of the $\sim$Milky Way-mass isolated elliptical galaxy \src. This has allowed
us to measure in fine detail both the total gravitating mass distribution and the baryon 
fraction out to large scales. In summary:

1. The relaxed X-ray morphology of \src, lack of radio
emission, and the good fits obtained from our 
hydrostatic models to the nontrivial temperature and density profile indicate that 
the gas must be close to hydrostatic.

2. {A constant M/L ratio model for the gravitating mass distribution
is ruled out in favour of a model including an
NFW dark matter halo at high significance ($\sim$20-$\sigma$)}.

3. {Assuming an NFW dark matter halo, we obtained unprecedented, tight constraints on both the 
virial mass and concentration for such a low-mass system 
(\mtwentyfive=$1.6\pm0.2 \times 10^{12}$\msun\ with systematic errors typically \ltsim 20\%; 
\mvir $=3.1^{+0.4}_{-0.3} \times 10^{12}$\msun).
This confirms that an isolated elliptical galaxy that is not at the centre of a massive
group can maintain a substantial dark matter halo.}

4. The total gas mass inferred within \rvir\ substantially exceeds the stellar mass,
with most of the gas lying outside $\sim$100~kpc. This supports theoretical
predictions that Milky Way-mass galaxies can host massive, quasi-hydrostatic hot gas halos.
It remains to be seen if less isolated galaxies can, similarly, maintain such a halo.

5. {Self-consistently extrapolated to \rvir, the baryon fraction is consistent with the Cosmological
value (0.17), indicating that it is possible even for a $\sim$Milky Way-mass galaxy
to be baryonically closed, at least if (relatively) isolated.}

6. Within both \rtwentyfive\ and \rfive, the gas fraction (and the baryon fraction 
in the case of \rfive) of \src\ are both consistent with
an extrapolation to low masses of the trends seen for groups and clusters.

7 After correcting for the gas fraction, the entropy profile is close to the self-similar predictions
of gravitational structure formation simulations, as observed in massive galaxy clusters.

8. Both the \chandra\ and \suzaku\ data reveal evidence of a strong heavy metal abundance gradient,
qualitatively similar to those observed in relaxed, massive galaxy groups and clusters.

\acknowledgements
We would like to thank Taotao Fang for insightful and helpful discussions.
We also benefitted greatly from discussions with H\'el\`ene Flohic, Erik Tollerud,
James Bullock and Pepi Fabbiano.
This research has made use of data obtained from the High Energy Astrophysics
Science Archive Research Center (HEASARC), provided by NASA's Goddard Space
Flight Center.
This research has also made use of the
NASA/IPAC Extragalactic Database (\ned)
which is operated by the Jet Propulsion Laboratory, California Institute of
Technology, under contract with NASA, and the HyperLEDA database
(http://leda.univ-lyon1.fr). We are grateful to the ACE/ SWEPAM instrument
team for making their data publicly available through the ACE Science Center.
PJH and DAB gratefully acknowledge partial support from NASA under Grant NNX10AD07G, 
issued through the office of Space Science Astrophysics Data Program. Partial
support for this work was also provided by NASA under Grant 
NNG04GE76G issued through the Office of Space Sciences Long-Term
Space Astrophysics Program, and by \chandra\ awards GO6-7071X and GO9-0092X,
issued by the Chandra X-Ray Center, which is
operated by the Smithsonian Astrophysical Observatory for and on behalf of NASA. 
We are also grateful for partial support from NASA-\suzaku\ grants 
NNX09AC71G and NNX10AH85G and NASA-\xmm\ grant NNX08AX74G.

\bibliographystyle{apj_hyper}
\bibliography{paper_bibliography.bib}

\end{document}